\documentclass[12pt]{article}
\usepackage{axodraw}
\usepackage{psfig,epsfig}
\usepackage{tabularx}
\def\21{$SU(2) \otimes U(1) $}
\def\ne{\hbox{$\nu_e$ }}
\def\nm{\hbox{$\nu_\mu$ }}
\def\nt{\hbox{$\nu_\tau$ }}
\def\mnt{\hbox{$m_{\nu_\tau}$ }}
\def\eq#1{{eq. (\ref{#1})}}

\evensidemargin=\oddsidemargin
\def\s#1{\tilde{#1}}
\newcommand{\mx}{\left[\begin{array}} 
\newcommand{\finmx}{\end{array}\right]} 
\newcommand{\mxp}{\left(\begin{array}} 
\newcommand{\finmxp}{\end{array}\right)} 
\def\beq{\begin{equation}}

\def\eeq{\end{equation}}
\def\bea{\begin{eqnarray}}
\def\eea{\end{eqnarray}}
\def\nn{\nonumber}

\def\fracs#1#2{{\hbox{$\frac{#1}{#2}$}}}
\def\mathbf#1{\hbox{\bf #1}}
\def\textrm#1{\hbox{#1}}
\def\bold#1{\hbox{\bf #1}}%\setbox0=\hbox{$#1$} 
\def\half{{\textstyle{1 \over 2}}} 
\def\third{{\textstyle{1 \over 3}}} 
\def\lsim{\raise0.3ex\hbox{$\;<$\kern-0.75em\raise-1.1ex\hbox{$\sim\;$}}}
\def\gsim{\raise0.3ex\hbox{$\;>$\kern-0.75em\raise-1.1ex\hbox{$\sim\;$}}}
\newcommand{\matriz}{\left[\begin{array}} 
\newcommand{\finmatriz}{\end{array}\right]} 
\newcommand {\ignore}[1]{}
\def\epj#1#2#3{         {Eur. Phys. J.}{\bf #1} (19#2) #3}

\def\np#1#2#3{           {Nucl. Phys. }{\bf #1} (19#2) #3}
\def\pl#1#2#3{           {Phys. Lett. }{\bf #1} (19#2) #3}
\def\pr#1#2#3{           {Phys. Rev. }{\bf #1} (19#2) #3}
\def\prep#1#2#3{         {Phys. Rep. }{\bf #1} (19#2) #3}
\def\prl#1#2#3{          {Phys. Rev. Lett. }{\bf #1} (19#2) #3}

\def\zp#1#2#3{           {Zeit. fur Physik }{\bf #1} (19#2) #3}

\def\n.c.#1#2#3{         {Nuovo Cim. }{\bf #1} (19#2) #3}
\def\r.n.c.#1#2#3{       {Riv. del Nuovo Cim. }{\bf #1} (19#2) #3}

\newcommand {\stbt} {\tilde{t}_1 \to b\,\tau }
\newcommand {\stch} [1] {\tilde{t}_1 \to c\,{\tilde{\chi}^0_{#1}} }
\newcommand {\stcha}{\tilde{t}_1 \to b \,{\tilde{\chi}^{+}_1} }
\def\rp{R-parity }
\def\nt{\hbox{$\nu_\tau$ }}
\newcommand{\chiz} [1] {\tilde{\chi}^{0}_{#1} }
\begin{document}
%\baselineskip=1.5\baselineskip
\begin{titlepage}
\begin{flushright}hep-ph/9908286\\ FTUV/99-25\\ 
IFIC/99-27\\ FSU-HEP-990712\\ UCCHEP/10-00\\ August 1999
\end{flushright}
\vspace*{5mm}
\begin{center} 
{\Large \bf Two-Body Decays of the Lightest Stop in Minimal Supergravity 
with and without R-Parity }\\[15mm]
{\large Marco A. D\'\i az${}^{1,3}$, Diego. A. Restrepo${}^2$, and 
Jos\'e. W. F. 
Valle${}^2$}\\
\hspace{3cm}\\
{\small ${}^1$Department of Physics, Florida State University,}\\ 
{\small Tallahassee, Florida 32306, USA}
\vskip.5cm
{\small ${}^2$Departamento de F\'\i sica Te\'orica, IFIC-CSIC, 
Universidad de Valencia}\\ 
{\small Burjassot, Valencia 46100, Spain}
\vskip.5cm
{\small ${}^3$Departamento de F\'\i sica, Universidad Cat\'olica de
Chile}\\ 
{\small Av. Vicu\~na Mackenna 4860, Santiago 6904411, Chile}
\hspace{3cm}\\
\end{center}
\vspace{5mm}
\begin{abstract}
  
  We study the decays of the lightest top squark in supergravity
  models with and without R-parity. Using the simplest model with an
  effective explicit bilinear breaking of R-parity and radiative
  electroweak symmetry breaking we show that, below the threshold for
  decays into charginos $\stcha$, the lightest stop decays mainly into
  third generation fermions, $\stbt$ instead of the R-parity
  conserving mode $\stch{1}$, even for tiny tau--neutrino mass values.
  Moreover we show that, even above the threshold for decays into
  charginos, the decay $\stbt$ may be dominant. We study the role
  played by the universality of the boundary conditions on the soft
  supersymmetry breaking terms.  This new decay mode $\stbt$ as well
  as the cascades originated by the conventional $\stch{1}$ decay
  followed by the \rp violating neutralino decays can provide new
  signatures for stop production at LEP and the Tevatron.
\end{abstract}

\end{titlepage}

\setcounter{page}{1}

\section{Introduction}

The {\sl Minimal Supersymmetric Standard Model (MSSM)}~\cite{mssm} or
its minimal supergravity (SUGRA) version~\cite{msugra} are by far the
most well studied realizations of supersymmetry. However, neither
gauge invariance nor supersymmetry requires the conservation of
R-parity. Indeed, there is considerable theoretical and
phenomenological interest in studying possible implications of
alternative scenarios~\cite{beyond} in which R-parity is
broken~\cite{aul,expl0,rpold,arca}.  This is especially so considering
the fact that it provides an appealing joint explanation of the solar
and atmospheric neutrino anomalies which has, in addition, the virtue
of being testable at future accelerators like the
LHC~\cite{Romao:1999up}. The violation of R-parity could arise
explicitly~\cite{expl} as a residual effect of some larger unified
theory~\cite{expl0}, or spontaneously, through nonzero vacuum
expectation values (vev's) for scalar neutrinos~\cite{aul,rpold,arca}.
In realistic spontaneous R-parity breaking models there is an \21
singlet sneutrino vev characterizing the scale of R-parity violation
\cite{MASIpot3,MASI,ROMA,ZR} which is set by the supersymmetry
breaking scale.

There are two generic cases of spontaneous R-parity breaking models to
consider.
In the absence of any additional gauge symmetry, these models lead to
the existence of a physical massless Nambu-Goldstone boson, called
majoron (J) which is {\sl the lightest SUSY particle}, massless and
therefore stable. It plays an important role  in making these models
fully consistent with astrophysics and cosmology~\cite{beyond} if one
wishes to contemplate the case of large breaking scales and heavy tau
neutrino.  If the majoron acquires a small mass due to explicit
breaking effects at the Planck scale then it may decay into electron
and muon neutrinos or photons, on very large time scales of
cosmological interest, playing a possible role as unstable dark
matter~\cite{KEV}.  
Alternatively, if lepton number is part of the gauge symmetry and \rp
is spontaneously broken then there is an additional gauge boson which
gets mass via the Higgs mechanism, and there is no physical Goldstone
boson~\cite{ZR}.  As in the standard case in \rp breaking models the
lightest SUSY particle (LSP) is in general a neutralino. However, it
now decays mostly into visible states, therefore diluting the missing
momentum signal and bringing in increased multiplicity events which
arise mainly from three-body decays such as
\beq
\label{vis}
\chiz{1} \to  f \bar{f} \nu,
\eeq
where $f$ denotes a charged fermion. The neutralino also has the
invisible
decay mode
\beq
\label{invi}
\chiz{1} \to  3 \nu.
\eeq
as well as
\beq
\label{invis}
\chiz{1} \to  \nu\,J,
\eeq
in the case the breaking of \rp is spontaneous~\cite{MASIpot3,MASI}.
This last decay conserves R-parity since the majoron has a large R-odd
singlet sneutrino component.

If R--parity is broken then supersymmetric (SUSY) particles need not
be produced only in pairs, and the lightest of them could decay.  The
effects of \rp violation can be large enough to be experimentally
observable.

In this paper we focus on the decay modes of the lightest top squark
in supergravity models where supersymmetry is realized with R-parity
violation. In such models the lightest stop could even be the lightest
supersymmetric particle and be produced at LEP. Neither $e^+ e^-$
collider data~\cite{LEPSEARCH} nor $p\bar{p}$ data from the Tevatron
\cite{D0} preclude this possibility. In contrast with
Ref.~\cite{stop1} here we focus on an effective model where the
breaking of R-parity is introduced through an explicit bilinear term
in the superpotential. This is substantially simpler than the full
majoron version of the model considered previously. In fact, this
bilinear model is not only especially simple theoretically, also its
phenomenological implications in collider physics can be studied in a
very systematic way. The bilinear model constitutes the simplest \rp
breaking model \cite{Diaz:1997xc} consistent with radiative
electroweak symmetry breaking, very much the same way as the minimal
\rp conserving supergravity models with universal soft SUSY breaking
terms~\cite{martin}, MSUGRA, for short.  As mentioned it also provides
an attractive joint explanation of the present neutrino anomalies
\cite{Romao:1999up}.

In order to discuss stop decays we also refine the work presented in
Ref.~\cite{HikKob,baer,porod,Boehm:1999tr} by giving, for the first
time, an exact numerical calculation for the FCNC process $\s t \to
c\, \s\chi_1^0$. We also compare the results obtained this way with
those one gets by adopting the usual one--step or leading logarithm
approximation in the RGE's. In contrast with the \rp conserving model
such an approximation would be rather poor for our purposes, since we
will be interested in comparing FCNC with \rp violating stop decay
modes (see section 5).
Moreover, in contrast to ref. \cite{stop1}, where the magnitude of the
stop -- charm -- neutralino coupling was a phenomenological parameter,
here we assume a minimal supergravity scheme with universality of soft
terms at the unification scale in which this coupling is induced
radiatively. As we will see this has important phenomenological
implications as for the behaviour of the \rp violating stop decays
with respect to $\tan\beta$. We calculate its magnitude using a set of
RGE's in which the running of the Yukawa couplings and soft breaking
terms is taken into account. Here we also provide the analysis of the
relationship of the \rp violating stop decays with the magnitude of
the tau neutrino mass. Motivated by the simplest oscillation
interpretation of the Super-Kamiokande atmospheric neutrino data, we
also generalize the treatment of the \rp violating decays by
explicitly considering the case of light tau--neutrino masses, not
previously discussed.

For definiteness and simplicity we focus on supersymmetric models
where the breaking of R-parity is parametrized explicitly through a
bilinear superpotential term of the type $\ell H_u$~\cite{epsi}. The
stop can have new decay modes such as
\beq
\label{btau}
\stbt
\eeq
due to mixing between charged leptons and charginos.  We show that
this decay may be dominant or at least comparable to the ordinary
R-parity conserving mode 
\beq
\label{chic}
\stch{1},
\eeq
where $\chiz{1}$ denotes the lightest neutralino.

The paper is organized as follows. The model and an analytical
analysis of the tree--level tau--neutrino in terms of SUGRA parameters
is described in section~\ref{section2}. The mass matrices are given in
section~\ref{section3} while in section 4 we present the top squark
decay widths in the minimal supergravity model with universal soft
SUSY breaking terms~\cite{martin}, MSUGRA, for short. The relevant
Feynman rules and the squark decay widths and branching ratios are
calculated in appendix~\ref{appendixb}. They are studied numerically
in section 5 and we present our conclusions in section 6.

\section{The Model}
\label{section2}

The supersymmetric Lagrangian is specified by the superpotential $W$
given by
\begin{equation}  
W=\varepsilon_{ab}\left[ 
 h_U^{ij}\widehat Q_i^a\widehat U_j\widehat H_2^b 
+h_D^{ij}\widehat Q_i^b\widehat D_j\widehat H_1^a 
+h_E^{ij}\widehat L_i^b\widehat R_j\widehat H_1^a 
-\mu\widehat H_1^a\widehat H_2^b 
\right] + \varepsilon_{ab}\epsilon_i\widehat L_i^a\widehat H_2^b\,,
\label{eq:Wsuppot} 
\end{equation}
where $i,j=1,2,3$ are generation indices, $a,b=1,2$ are $SU(2)$
indices, and $\varepsilon$ is a completely antisymmetric $2\times2$
matrix, with $\varepsilon_{12}=1$. The symbol ``hat'' over each letter
indicates a superfield, with $\widehat Q_i$, $\widehat L_i$, $\widehat
H_1$, and $\widehat H_2$ being $SU(2)$ doublets with hypercharges
$\third$, $-1$, $-1$, and $1$ respectively, and $\widehat U$,
$\widehat D$, and $\widehat R$ being $SU(2)$ singlets with
hypercharges $-{\textstyle{4\over 3}}$, ${\textstyle{2\over 3}}$, and
$2$ respectively. The couplings $h_U$, $h_D$ and $h_E$ are $3\times 3$
Yukawa matrices, and $\mu$ and $\epsilon_i$ are parameters with units
of mass.
 
Supersymmetry breaking is parametrized by the standard set of soft
supersymmetry breaking terms 
\begin{eqnarray} 
V_{soft}&=& 
M_Q^{ij2}\widetilde Q^{a*}_i\widetilde Q^a_j+M_U^{ij2} 
\widetilde U^*_i\widetilde U_j+M_D^{ij2}\widetilde D^*_i 
\widetilde D_j+M_L^{ij2}\widetilde L^{a*}_i\widetilde L^a_j+ 
M_R^{ij2}\widetilde R^*_i\widetilde R_j \nonumber\\ 
&&\!\!\!\!+m_{H_1}^2 H^{a*}_1 H^a_1+m_{H_2}^2 H^{a*}_2 H^a_2\nn\\
&&\!\!\!\!- \left[\half M_3\lambda_3\lambda_3+\half M\lambda_2\lambda_2 
+\half M'\lambda_1\lambda_1+h.c.\right] 
\nn\\ 
&&\!\!\!\!+\varepsilon_{ab}\left[ 
A_U^{ij}h_U^{ij}\widetilde Q_i^a\widetilde U_j H_2^b 
+A_D^{ij}h_D^{ij}\widetilde Q_i^b\widetilde D_j H_1^a 
+A_E^{ij}h_E^{ij}\widetilde L_i^b\widetilde R_j H_1^a\right.
\nn\\ 
&&\!\!\!\!\left.-B\mu H_1^a H_2^b+B_i\epsilon_i\widetilde L_i^a
H_2^b\right] 
\,,\label{eq:Vsoft}
\end{eqnarray} 

For definiteness and simplicity we assume only R-parity Violation
(RPV) in the third generation, neglecting the effects of RPV on the
two first families, adopting the superpotential
\cite{moreBRpV,otros,javi}
\begin{equation} 
W=h_t\widehat Q_3\widehat U_3\widehat H_2
 +h_b\widehat Q_3\widehat D_3\widehat H_1
 +h_{\tau}\widehat L_3\widehat R_3\widehat H_1
 -\mu\widehat H_1\widehat H_2
 +\epsilon_3\widehat L_3\widehat H_2
\label{eq:Wbil}
\end{equation}
to describe the R--Parity violating violating decay mode $\s t_1 \to
b\, \tau$.  In this case we will omit the labels $i,j$ in the soft
breaking terms given above.  Note that the bilinear term $\epsilon_3$
can not be rotated away, since the rotation that eliminates it
reintroduces an R--Parity violating trilinear term, as well as a
sneutrino vacuum expectation value. Notice that, in contrast with ref.
\cite{Romao:1999up} where the doublet sneutrino vev in the bilinear
model is much more loosely constrained, 
%%% as 
in this case  %this is the new
it is not subject to
constraints from astrophysics.

Note, in contrast, that in order to describe Flavour Changing Neutral
Current (FCNC) effects such as the R--Parity conserving process $\s
t_1 \to c\,\s\chi_1^0$ we need the three generations of quarks.

The above model can be described in various equivalent bases, for
example
\begin{enumerate}
\item
one in which bilinear term and sneutrino vev are non-zero, $\epsilon_3^I
\neq 0$ and $v_3^I \neq 0$~\cite{Diaz:1997xc,beyond}
\item
one in which trilinear 
%%%
term~\footnote{In the one generation case there is only one
trilinear RPV term in the superpotential written in our notation as
$\lambda_3\widehat Q_3\widehat D_3\widehat L_3$ } %this is the new 
and sneutrino vev are non-zero, $\lambda_3^{II} \neq 0$ and $v^{II}_3 \neq
0$~\cite{Borzumati:1999th}
\item
the vev-less basis in which $\epsilon_3^{III}$ and
$\lambda_3^{III}$ are non-zero but
$v_3^{III}=0$~\cite{Grossman:1998py,Bisset:1999nw}
\end{enumerate}
where the R-parity violating parameters can be expressed in terms of
dimension-less basis-independent alignment parameters $ \sin \xi$, $
\sin \xi'$ and $ \sin \xi''$~\cite{javi,sacha} ($X=I,II$ or $III$) as
follows:
\begin{eqnarray}
\label{supxi}
\sin\xi&=&\frac{\epsilon_3^Xv_d^X+\mu^X v_3^X}{\mu'{v_d'}}=
\frac{v_3^{II}}{{v_d'}}=\frac{\epsilon_3^{III}}{\mu'}\\
\label{supxip}
\sin\xi'&=&\frac{\mu^X \lambda_3^X+\epsilon_3^Xh_b^X}{\mu'h_b'}=
\frac{\lambda_3^{II}}{h_b'}=\frac{\epsilon_3^I}{\mu'}\\
\label{supxipp}
\sin\xi''&=&\frac{-v_d^X\lambda_3^X+v_3^Xh_b^X}{{v_d'}h_b'}=
\frac{v_3^I}{{v_d'}}=-\frac{\lambda_3^{III}}{h_b'}
\end{eqnarray}
where
\begin{equation} 
h_b'=\sqrt{{h_b^X}^2+{\lambda_3^X}^2}\qquad
\mu'=\sqrt{{\mu^X}^2+{\epsilon_3^X}^2}\qquad
{v_d'}=\sqrt{{v_d^X}^2+{v_3^X}^2}, \quad X=I,II,\hbox{ or }III
\label{hbmuvd}
\end{equation}
%

%%%
Note that, in the notation of eqs.~(\ref{supxi})--(\ref{supxipp}), the
parameters $\epsilon_3$ and $\mu$ appearing in eq.~(\ref{eq:Wbil})
should bear the superscript I.

Of these parameters only two are independent because they satisfy
\begin{equation}
\sin\xi''=\cos\xi'\sin\xi-\sin\xi'\cos\xi
\label{sinxippp}
\end{equation}

In the limit when the \rp violating parameters vanish one recovers the
MSSM. From now on we will work in the $\lambda_3^I=0$--basis, unless
otherwise stated.  As a result we will omit the label $I$ in all the
parameters associated with this basis. We also will drop out the prime
in $h_b$. One of the advantages in working in this basis is that the
RGE's evolution does not induce the trilinear \rp violating terms
neither in the superpotential nor in the scalar potential if at the
beginning we impose universality~\cite{javi}.

It is convenient to introduce the following notation in spherical
coordinates for the vacuum expectation values (vev):
\begin{eqnarray} 
v_d&=&v\sin\theta\cos\beta\cr 
v_u&=&v\sin\theta\sin\beta\cr 
v_3&=&v\cos\theta 
\label{eq:vevs} 
\end{eqnarray} 
which preserves the standard MSSM definition $\tan\beta=v_u/v_d$. In
the MSSM limit, where $\epsilon_3=v_3=0$, the angle $\theta$ is equal
to $\pi/2$. This makes sense in the $\lambda_3^I=0$--basis where the
usual MSSM relation
\begin{equation}
m_b=%&=&\frac1{\sqrt2}\left(h_bv_d+\lambda_3v_3\right)\\%&=&
\frac1{\sqrt2}h_bv_d\\
%&=&\frac1{\sqrt2}h_b'{v_d'}\cos\xi''
\label{mb}
\end{equation}
holds. 

In this model the presence of RPV induces a mass for the tau--neutrino
at the tree level~\cite{rpold,arca}, as well as radiative masses to
the the \ne and \nm. As already mentioned it is sufficient for our
present discussion of stop decays to keep only the tau--neutrino.  

In order to study the \nt mass it is convenient to have an analytical
expression for \mnt in this limit. The tree level tau--neutrino mass
may be expressed as~\cite{expl,expl0,javi}~--~\cite{anomaly}
\begin{equation}
m_{\nu_{\tau}} \approx 
-\frac{(g^2M'+g'^2M){\mu'}^2}{
4 M M'{\mu'}^2-2(g^2M'+g'^2M){\mu'}v_u{v_d'} \cos\xi}{v_d'}^2\sin^2\xi
\label{mNeutrinoApp}
\end{equation}
in terms of basis-independent parameters $\mu'$, $v_d'$ and $\sin \xi$
defined in Eqs~(\ref{hbmuvd}) and (\ref{supxi}).
The second term in the denominator may be neglected if $M,\mu\gsim
m_Z$, as often happens in minimal supergravity models with universal
soft SUSY breaking terms~\cite{martin}.  Thus one may obtain an
estimate of the neutrino mass by keeping only the first term in the
denominator.
\begin{equation}
m_{\nu_{\tau}}\approx\frac{g^2}{2M}{v_d'}^2\sin^2\xi
\label{mNeutrinoAppp}
\end{equation}
where we have used $M'=Mg'^2/g^2$.  For $\sin \xi \approx 1$ one can
easily check that \mnt could be as large as the experimental upper
bound of 18 Mev~\cite{ubmnt}. However in MSUGRA models one may obtain
naturally small $\sin \xi $ values, calculable from the RGE evolution
from the unification scale down to the weak scale. Indeed, using the
minimization equations $\sin\xi$ can be written in terms $\Delta
m^2=m^2_{H_1}-m^2_{L_3}$ and $\Delta B=B_3-B$~\cite{rptalks} as
\begin{equation}
\sin\xi=-\cos\xi'\sin\xi'
\left(\cos\xi\frac{\Delta m^2}{{m'}_{\tilde\nu^0_{\tau}}^2}+
\frac{v_2}{v_d'}\frac{\mu'\Delta B}{{m'}_{\tilde\nu^0_{\tau}}^2}\right)
\label{sinxi}
\end{equation}
One may give a simplified approximate analytical expression for the
tau--neutrino mass in this model by solving the renormalization group
equations for the soft mass parameters $m^2_{H_1}$, $m^2_{L_3}$, $B$,
and $B_3$ in the one--step approximation.  This gives ~\cite{rptalks}
\begin{eqnarray}
\sin\xi\left|_{\Delta m^2}\right.&\approx&-\cos\xi'\sin\xi'\cos\xi
h_b^2\left[\frac{m_{H_1}^2+M_Q^2+M_D^2+A_D^2}
{{m'}_{\tilde\nu^0_{\tau}}^2}\right]\left(\frac3{8\pi^2}
\ln\frac{M_{U}}{m_t}\right)\nn\\
&\sim&-\cos\xi'\sin\xi'\cos\xi
h_b^2\left(\frac3{8\pi^2}\ln\frac{M_{U}}{m_t}\right)
\label{sinxidm}
\end{eqnarray}
and
\begin{equation}
\sin\xi\left|_{\Delta B}\right.\approx\cos\xi'\sin\xi'
\tan\beta'
h_b^2\left[\frac{\mu'A_D}{{m'}_{\tilde\nu^0_{\tau}}^2}\right]
\left(\frac3{8\pi^2}\ln\frac{M_{U}}{m_t}\right)
\label{sinxidb}
\end{equation} 
where we have denoted by the symbols $\sin\xi|_{\Delta m^2}$ and
$\sin\xi|_{\Delta B}$ the two terms contributing to $\sin\xi$ in
eq.~(\ref{sinxi}). Using these expressions and assuming no strong
cancellation between these terms one finds that the minimum neutrino
mass is controlled by the $\sin\xi|_{\Delta m^2}$. As a result one
finds,
\begin{equation}
m_{\nu_{\tau}}\left|_{\mathrm{min}}\right.\sim
\frac{g^2m_b^2}{M}\left(\sin^2\xi'
h_b^2\right)\left(\frac3{8\pi^2}
\ln\frac{M_{U}}{m_t}\right)^2
\label{mNeutrinods}
\end{equation}

The above approximate analytical form of the tau-neutrino mass is
useful, as we will see later (e.g. \eq{tausa}) in order to display
explicitly the degree of correlation between the \rp violating decays,
such as $\tilde t_1\to b\tau$, with the tau-neutrino mass.

The minimum value for $\sin \xi' h_b$ is determined by the value $\sin
\xi'$ and that of $\tan \beta$. For $\sin \xi' \sim 1$ and relatively
small 
$\tan \beta$ so that $h_t$ is perturbative, one has
\begin{equation}
m_{\nu_{\tau}}\gsim 10 \hbox{KeV}
\label{1mev}
\end{equation}
for $M \sim 1$ TeV. In order to get smaller \nt masses one needs to
suppress $\sin^2 \xi'$ additionally, for example to reach one
electron-volt the required R-parity violating parameters are given in
Table~\ref{rpvev}.  These order-of-magnitude estimates are given in
terms of the basis--independent angles $\xi$ and $\xi'$, and in the
relevant parameters for the three bases defined before. 

Note that whenever the parameter has two values, the first correspond
to $\tan\beta=2$ (the lower perturbativity limit) and the second to
$\tan\beta=35$. In Table~\ref{rpvev}, $\sin\xi$ was estimated from
eq.~(\ref{mNeutrinoAppp}) and $\sin\xi'$ from eq.~(\ref{mNeutrinods}).

Note that the RGE-induced suppression depends basically in the $h_b^2$
factor in eq.~(\ref{sinxidm}) which is $\sim 10^{-3}$ ($\sim 1$) for
small (large) $\tan\beta$. As a result the bigger the value of
$\tan\beta$, the smaller $\sin\xi'$ will have to be for a fixed tau
neutrino mass. The RPV parameters in the several bases were estimated
from Eqs.~(\ref{supxi}--\ref{supxipp}) and~(\ref{sinxippp}). 

\begin{table}
\begin{center}
\begin{tabular}{|c|c|c|c|c|c|c|c|c|c|c|c|c|}\hline
\multicolumn{4}{|c|}{basis--independent}&
\multicolumn{4}{|c|}{Basis I: $\lambda_3^I=0$}&
\multicolumn{2}{c|}{Basis II: $\epsilon_3^{II}=0$}&
\multicolumn{3}{c|}{Basis III: $v_3^{III}=0$ }\\ \hline
\multicolumn{2}{|c|}{$\sin\xi$}&
\multicolumn{2}{c|}{$\sin\xi'$}&
\multicolumn{2}{|c|}{$\!\!\epsilon_3(\hbox{GeV})\!\!$}&
\multicolumn{2}{c|}{$\!\!v_3(\hbox{GeV})\!\!$}&
$\lambda_3^{II}$&$\!\!v_3^{II}(\hbox{GeV})\!\!$&$\lambda_3^{III}$&
\multicolumn{2}{c|}{$\!\!\epsilon_3^{III}(\hbox{GeV})\!\!$}\\ \hline
$\!\!10^{-5}\!\!$&$\!\!10^{-4}\!\!$&$\!\!10^{-2}\!\!$&
$\!\!10^{-4}\!\!$&1&$\!\!10^{-2}\!\!$&1&$\!\!10^{-3}\!\!$&
$\!\!10^{-4}\!\!$&$\!\!10^{-3}\!\!$&$\!\!10^{-4}\!\!$&
$\!\!10^{-3}\!\!$&$\!\!10^{-2}\!\!$\\ \hline
\end{tabular}
\end{center}
\caption{\small
  Estimated magnitude of R-parity violating parameters required for a
  tau--neutrino mass in the eV range, without requiring cancellation
  in $\sin\xi$ in the three bases defined before. }
\label{rpvev}
\end{table}

In \eq{mNeutrinods} we have neglected $\Delta B$ contribution with
respect to the one coming from $\Delta m^2$. It is possible, however,
that the $\Delta B$ term may be sizeable. In the large $\Delta B$ case
then it may cancel the $\Delta m^2$ contribution in $\sin\xi$, leading
to an additionally suppressed neutrino mass. As we will see, however,
in SUGRA models with universal soft terms at the unification scale
($\epsilon$SUGRA for short) we do not need any substantial
cancellation in order to obtain \nt masses below the electron-volt
scale.

\section{Squark Mass Matrices}
\label{section3}

The up and down-type squark mass matrices of our model have already
been given previously in Ref.~\cite{Diaz:1997xc}. Here we generalize
those to the three-generation case. The mass matrix of the up squark
sector follows from the quadratic terms in the scalar potential
\begin{equation} 
V_{quadratic}=\mx{cc}
\s {\bold{u}}_L^\dagger&\s{\bold{ u}}_R^\dagger
\finmx 
{\s M_U^2}\mx{c}
\s{\bold{ u}}_L\\
\s{\bold{ u}}_R
\finmx+\cdots
\end{equation} 
given by 
\beq
{\s M_U^2}=\mx{cc}
{M_Q^2}+\frac12v_u^2{h_U}{h_U}^\dagger+\Delta_{UL}&
\frac1{\sqrt2}v_u{A_U^h}-\frac1{\sqrt2}(\mu v_d-\epsilon_3v_3)
{h_U}\\
\frac1{\sqrt2}v_u{A_U^h}^\dagger-\frac1{\sqrt2}(\mu v_d-\epsilon_3v_3)
{h_U}^\dagger &
{M_U^2}+\frac12v_u^2{h_U}^\dagger{h_U}+\Delta_{UR}
\finmx
\eeq
where $\Delta_{UL}=\frac18\big(g^2-\frac13{g'}^2\big)\big(v_d^2-v_u^2
+v_3^2\big)\bold{1}$ and $\Delta_{UR}=\frac16
{g'}^2(v_d^2-v_u^2+v_3^2)\bold{1}$ are the splitting in the squark
mass spectrum produced by electro-weak symmetry breaking, and
${A_U^h}_{ij}\equiv A_U^{ij}h_U^{ij}$. The eigenvalues of ${\s M_U^2}$
are
\begin{equation}
\textrm{diag}\,\{m_{\s u_1},m_{\s u_2},\ldots,m_{\s u_6}\}=\mx{cc}
\Gamma_{UL}& 
\Gamma_{UR}
\finmx
{\s M_U^2}
\mx{c}
\Gamma_{UL}^\dagger \\
\Gamma_{UR}^\dagger
\finmx
\end{equation}
This way the six weak-eigenstate fields $\s u_{iL}$ and $\s u_{iR}$
($i=1,2,3$) combine into six up-type mass eigenstate squarks $\s u_k$
as follows: $\s u_{iL}=\Gamma^{\dagger ik}_{UL}\s u_k=\Gamma^{*
ki}_{UL}\s u_k$, $\s u_{iR}=\Gamma^{\dagger ik}_{UR}\s u_k=\Gamma^{*
ki}_{UR}\s u_k$.

For completeness, we also give the mass matrix of the down squark
sector. The quadratic scalar potential includes
\begin{equation} 
V_{quadratic}=\mx{cc}
\s{\bold{d}}_L^*&\s{\bold{d}}_R^*
\finmx 
{\s M_D^2}\mx{c}
\s{\bold{d}}_L\\
\s{\bold{d}}_R
\finmx+\cdots
\end{equation} 
given by 
\beq
{\s M_D^2}=\mx{cc}
{M_Q^2}+\frac12v_d^2{h_D}{h_D}^\dagger+\Delta_{DL}&
\frac1{\sqrt2}v_d{A_D^h}-\frac1{\sqrt2}\mu v_u
{h_D}\\
\frac1{\sqrt2}v_d{A_D^h}^\dagger -\frac1{\sqrt2}\mu v_u
{h_D}^\dagger  &
{M_D^2}+\frac12v_d^2{h_D}^\dagger{h_D}+\Delta_{DR}
\finmx
\eeq
where $\Delta_{DL}=-\frac18\big(g^2+\frac13{g'}^2\big)\big(v_d^2-v_u^2
+v_3^2\big)\bold{1}$, $\Delta_{DR}=-\frac1{12}
{g'}^2(v_d^2-v_u^2+v_3^2)\bold{1}$, and
${A_D^h}_{ij}\equiv A_D^{ij}h_D^{ij}$. The eigenvalues of ${\s M_D^2}$
are 
\begin{equation}
\textrm{diag}\,\{m_{\s d_1},m_{\s d_2},\ldots,m_{d_6}\}=\mx{cc}
\Gamma_{DL}& 
\Gamma_{DR}
\finmx
{\s M_D^2}
\mx{c}
\Gamma_{DL}^\dagger \\
\Gamma_{DR}^\dagger
\finmx
\end{equation}
One is left with six mass-eigenstate down squarks fields $\s d_{k}$
related to $\s d_{iL}$ and $\s d_{iR}$ fields as follows: $\s
d_{iL}=\Gamma^{\dagger ik}_{DL}\s d_k=\Gamma^{* ki}_{DL}\s d_k$, $\s
d_{iR}=\Gamma^{\dagger ik}_{DR}\s d_k=\Gamma^{* ki}_{UR}\s d_k$.

For the Higgs-slepton part of the quadratic scalar potential in the
one generation case of the Bilinear R-parity Violating (BRpV) model,
see refs.~\cite{v3cha} and \cite{epsi}.

Of particular interest to us is the chargino/tau mass matrix. For our
present purposes it is sufficient to have the form of this matrix for
one generation, which is given by
\begin{equation}
{\bf M_C}=\left[\matrix{
M & {\textstyle{1\over{\sqrt{2}}}}gv_2 & 0 \cr
{\textstyle{1\over{\sqrt{2}}}}gv_d & \mu & 
-{\textstyle{1\over{\sqrt{2}}}}h_{\tau}v_3 \cr
{\textstyle{1\over{\sqrt{2}}}}gv_3 & -\epsilon_3 &
{\textstyle{1\over{\sqrt{2}}}}h_{\tau}v_d}\right]
\label{eq:ChaM6x6}
\end{equation}
This form is common to all models with spontaneous breaking of
R-parity, as well as in the simplest truncation of these models
provided by the BRpV model considered here.  We note that the chargino
sector decouples from the tau sector in the limit $\epsilon_3=v_3=0$.
As in the MSSM, the chargino mass matrix is diagonalized by two
rotation matrices $\bf U$ and $\bf V$
\begin{equation}
{\bf U}^*{\bf M_C}{\bf V}^{-1}=\left[\matrix{
m_{\s\chi^{\pm}_1} & 0 & 0 \cr
0 & m_{\s\chi^{\pm}_2} & 0 \cr
0 & 0 & m_{\tau}}\right]\,.
\label{eq:ChaM3x3}
\end{equation}
The lightest eigenstate of this mass matrix must be the tau lepton
($\tau^{\pm}$) and so the mass is constrained to be
$1.77705^{+0.29}_{-0.26}$ GeV.  To obtain this the tau Yukawa coupling
becomes a function of the parameters in the mass matrix, and the full
expression is given in~\cite{v3cha}. The composition of the tau is
given by
\begin{equation}
\tau^+_R=V_{3j}\psi_j^+, \qquad \tau^-_L=U_{3j}\psi_j^- 
\label{taucomp}
\end{equation}
where $\psi^{+T}=(-i\lambda^+,\widetilde H_2^1,\tau_R^{0+})$ and
$\psi^{-T}=(-i\lambda^-,\widetilde H_1^2,\tau_L^{0-})$. The
two-component Weyl spinors $\tau_R^{0-}$ and $\tau_L^{0+}$ are weak
eigenstates, while $\tau^+_R$ and $\tau^-_L$ are the mass eigenstates.
It follows easily from eq.~(\ref{eq:ChaM3x3}) that the matrix
${\bf{M_CM_C^T}}$ is diagonalized by ${\bf U}$ and the matrix
${\bf{M_C^TM_C}}$ is diagonalized by ${\bf V}$.

The soft SUSY breaking parameters at the electroweak scale needed for
the evaluation of the mass matrices and couplings are calculated by
solving the renormalization group equations (RGE's) of the MSSM and
imposing the radiative electroweak symmetry breaking condition. From
the measured quark masses, CKM matrix elements and $\tan\beta$ we
first solve one-loop RGE's for the gauge and Yukawa couplings to
calculate their corresponding values at the unification scale.
Assuming now universal soft supersymmetry breaking boundary
conditions, we evolve downwards the RGE's for all MSSM parameters,
including full three-generation mixing in the RGE's for Yukawa
coupling constants, as well as soft SUSY breaking parameters. Next, we
evaluate the Higgs potential at the $m_t$ scale including the one-loop
corrections induced by the Yukawa coupling constants of the third
generation. The radiative electroweak symmetry breaking requirement
fixes the magnitude of the SUSY Higgs mass parameter $\mu$ and the
soft SUSY breaking parameters $B$ and $B_3$.  Notice that due to the
third minimization condition one can solve for $v_3$ as a function of
$\epsilon_3$.  At this point, all RPV parameters at the electroweak
scale are determined as functions of the input parameters
$(\tan\beta,\, m_0,\, A_0,m_{1/2}, \hbox{sign}(\mu), \epsilon_3)$.
Iteration is required because $\mu$ and $\epsilon_3$ are inputs to
evaluate the loop-corrected minimum.  Having determined all parameters
at the electroweak scale, we obtain the masses and the mixings of all
the SUSY particles by diagonalizing the corresponding mass matrices.
At this stage we also choose $\epsilon_3$ in order to get a
sufficiently light tau--neutrino.

We scan the soft SUSY breaking parameter space in the range 
\begin{eqnarray}
&m_0&\le 700 \,\hbox{GeV}\nn \\
100 \:\hbox{GeV}<&m_{1/2}&\le 400 \,\hbox{GeV}\nn \\ 
\label{scan0}
&|A_0|&\le 1000 \,\hbox{GeV}, \\
&|\epsilon_3|&<200 \,\hbox{GeV}\nn \\
1.8<&\tan\beta&<60 \nn
\end{eqnarray}
the previous range on $\tan\beta$ guarantee that both $h_t$ and $h_b$
will be perturbative. For the CKM matrix, we use the Particle Data
Group convention~\cite{Caso:1998tx}, taking $V_{us}=0.2205$,
$V_{cb}=0.041$, $|V_{ub}/V_{cb}|=0.08$ and neglecting CP violation,
i.e. $\delta=0$. Notice that here we scan over a much larger range for
epsilon than used in ref. \cite{stop1}.

\begin{figure}
\centerline{\protect\hbox{
\psfig{file=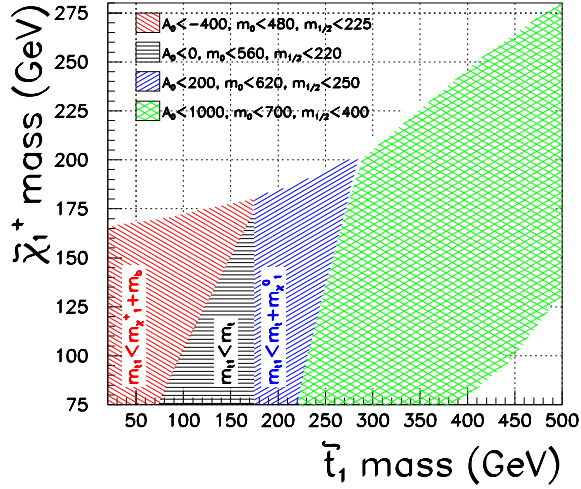,height=7cm}}}
\caption{\small 
  Kinematical regions in the $m_{{\tilde
      t}_1}$--$m_{{\tilde\chi}_1^+}$ plane.  From left to right:
  Region I $m_{\tilde t_1} < m_{{\tilde\chi}_1^+}+m_b$; Region II
  $m_{{\tilde\chi}_1^+}+m_b<m_{{\tilde t}_1}<m_t$; Region III
  $m_t<m_{{\tilde t}_1}<m_{{\tilde\chi}_1^0}+m_t$; and region IV
  $m_{{\tilde t}_1}>m_{{\tilde\chi}_1^0}+m_t$ }
\label{par-sp}
\end{figure}

The resulting region of lightest stop and chargino masses is displayed
in Fig.~\ref{par-sp}. Neglecting the three-body decays, we find that
in Region I of the $m_{{\tilde t}_1}$--$m_{{\tilde\chi}_1^+}$ plane,
$BR(\s t_1 \to c\,\s\chi_1^0)+ BR(\s t_1 \to b\,\tau)\approx 1$. In
Region II $BR(\s t_1 \to b\,\tau)+BR(\s t_1 \to b\,\s\chi_i^+)\approx 1$
(i=1,2).  In Region III $BR(\s t_1 \to b\,\tau)+BR(\s t_1 \to
b+\s\chi_i^+)+BR(\s t_1 \to t\,\nu_\tau)\approx 1$ (i=1,2), while in
region IV $BR(\s t_1 \to b\,\tau)+BR(\s t_1 \to b\,\s\chi_i^+)+BR(\s
t_1 \to t\,\nu_\tau)+BR(\s t_1  \to t\,\s\chi_j^0) \approx 1$
$(j=1,\ldots,4)$. Note that in each region the exact equality to 1 is
reached when the FCNC processes are fully included.

In Appendix~\ref{appendixb} we give the Feynman rules for all vertices
involving squarks, quarks, charginos and neutralinos, as well as the
two--body squark decay-widths, for squarks of all three generations.
These equations reduce to the expressions found in Ref.~\cite{BMP}
provided one identifies $\Gamma_{UL}^{33}=\cos\theta_{\s t}$ and
$\Gamma_{UR}^{33}=\sin\theta_{\s t}$.  They also generalize the
results for the BRpV model to the three-generation case.

\section{Lightest Stop Two-Body Decays in MSUGRA}

In an R--parity conserving supergravity theory the main $\s t_1$ decay
channel expected in region I of Fig.~\ref{par-sp} is the loop--induced
and flavour--changing $\s t_1 \to c\,\s\chi_1^0$
~\cite{HikKob,baer,porod}.  As is well-known, the FCNC processes in
the MSSM in general involve a very large number of input parameters.
For this reason, following common practice, we prefer to perform the
phenomenological study of flavour changing processes in the framework
of a supergravity theory with universal supersymmetry breaking.  The
simplest description of FCNC processes in \rp conserving minimal SUGRA
models uses the so-called one-step approximation.  Here we start by
reproducing the standard calculation for $\s t_1 \to c\,\s\chi_1^0$ as
in~\cite{HikKob}. To do this consider only the effect of the third
generation Yukawa coupling. From our general eq.~(\ref{tause}) we have
for $\s t_1= \s u_l$
\begin{equation}
\Gamma(\s t_1 \to c\,\s\chi_1^0)
\approx \frac{g^2}{8\pi} \left(\Gamma_{UL13}\right)^2
\left[
\fracs23 \sin\theta_W
N'_{11}+\left(\fracs12-\fracs23\sin^2\theta_W\right)
\frac{N'_{12}}{\cos\theta_W}\right]^2
m_{\s t_1}
\left(1-\frac{m_{\s \chi_1^0}^2}{m_{\s t_1}^2}\right)^2
\label{gtn2cos}
\end{equation}
with
\begin{equation}
\Gamma_{UL13} =
\frac{\Delta_L\cos\theta_{\s t}-\Delta_R\sin\theta_{\s t}}
{m_{\s c_L}^2-m_{\s t_L}^2}
\end{equation}
In the one--step approximation $\Delta_L, \Delta_R$ are given by
\begin{eqnarray}
\Delta_L&=&(\s M_U^2)_{23}\approx (M_Q^2)_{23}\approx-
\frac{t_U}{16\pi^2}K_{cb}K_{tb}h_b^2(M_Q^2+M_D^2+m_{H_1}^2+A_b^2)\\
\Delta_R&=&(\s M_U^2)_{26}\approx (A_U^h)_{23}\approx-
\frac{t_U}{16\pi^2}K_{cb}K_{tb}h_b^2m_t(A_b+\frac12A_t)
\end{eqnarray}
with $t_U=\ln(M_G/m_t)$. So, in the one--step approximation we have
\small{
\begin{equation}
\label{gtn2coshb}
\Gamma(\s t_1 \to c\,\s\chi_1^0)\approx F h_b^4(
\delta_{m_0^2}\cos\theta_{\s t}-\delta_A\sin\theta_{\s t})^2
\left[
\frac{\sqrt2}6(\tan\theta_W\,N_{11}+3N_{12})\right]^2
m_{\s t_1} \left(1-\frac{m_{\s \chi_1^0}^2}{m_{\s t_1}^2}\right)^2
\end{equation}
}
where the pre--factor $F = \frac{g^2}{16\pi}\left(
  \frac{t_U}{16\pi^2}K_{cb}K_{tb}\right)^2 \sim 6 \times 10^{-7}$ and
the parameter $\delta_{m_0^2}$ is given by
\begin{equation}
\delta_{m_0^2}=\frac{M_Q^2+M_D^2+m_{H_1}^2+A_b^2}{m_{\s c_L}^2-m_{\s
t_L}^2}
\sim 1
\end{equation}
is basically independent of the initial conditions due to the $m_0$
dependence both in the numerator as in the denominator and
\begin{equation}
\delta_A=\frac{m_t(A_b+\frac12A_t)}{m_{\s c_L}^2-m_{\s t_L}^2}
\end{equation}

\begin{figure}
\centerline{\protect\hbox{
\psfig{file=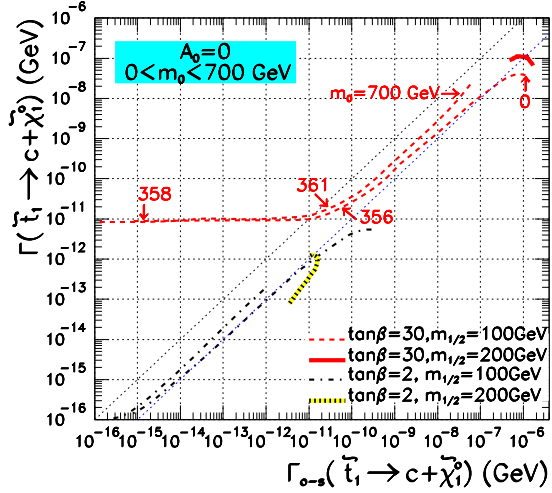,height=7cm}}}
\caption{\small 
  Comparison between the exact numerical calculation (ordinate) and
  the one--step approximation (abscissa) for the $\s t_1 \to
  c\,\s\chi_1^0$ decay width for various values of $\tan\beta$ and
  $m_{1/2}$ with $A_0=0$ and $m_0$ varying in the indicated range. The
  dotted left diagonal line would signify the equality between the
  estimates, while the right diagonal line would indicate one order of
  magnitude difference. Results of both estimates indicated in the
  lower right legend. More details are found in the text.}
\label{exact-os}
\end{figure}

Note however, that the one-step approximation includes only the third
generation Yukawa couplings and neglects the running of the soft
breaking terms~\cite{HikKob,baer,porod,Boehm:1999tr}.  Such an
approximation is rather poor for our purposes, since we will be
interested in comparing with \rp violating decay modes (see the next
section). In order to have an accurate calculation of the respective
branching ratios we need to go beyond the one-step approximation.  We
therefore use a exact numerical calculation for the FCNC process $\s t
\to c\, \s\chi_1^0$ in which the running of the Yukawa couplings and
soft breaking terms is taken into account.  First we have checked that
indeed the effect of the Yukawas from the two first generations is
negligible. However the same is not true for the running of the soft
breaking terms.  As can be seen from Figure~\ref{exact-os} the range
of variation that we obtain from the numerical solution is
\begin{equation}
\Gamma(\s t_1 \to c\,\s\chi_1^0)\sim(10^{-16}\hbox{ --
}10^{-6})\hbox{GeV}
\end{equation}
depending on the assumed value of $m_{1/2}$ and $\tan\beta$. In this
figure we have compared the decay width obtained from
eq.~(\ref{tause}) with the approximate formulae in eq.~(\ref{gtn2cos})
for two fixed values of $m_{1/2}$, $\tan\beta$ and taking $A_0=0$. The
approximate formulae only reproduce well the numerical result for the
academic case of no SUSY breaking gaugino mass, $m_{1/2}=0$. For the
more realistic case $m_{1/2}>100\,$GeV, the exact solution is usually
one decade smaller than the approximate one. In the one-step
approximation $\Gamma(\s t_1 \to c\,\s\chi_1^0)$ can be arbitrarily
small if the two terms $\delta_{m_0^2} \cos\theta_{\s t}$ and
$\delta_A\sin\theta_{\s t}$ in eq.~(\ref{gtn2coshb}) cancel. This
behaviour can be illustrated in Figure~\ref{exact-os} by the dashed
line labelled 358, which corresponds to $m_0 = 358$ GeV.  One sees
clearly that while the approximate solution goes to zero, the
numerical one reaches a minimum value around $10^{-11}$ GeV.  The
wrong behaviour of the approximate solution indicates that the
$\delta_A$ depends strongly on the scale.  For example, the RGE for
$A_b$ is very sensitive on $m_{1/2}$ and $\tan\beta$ and in the
one-step approximation there is no explicit dependence on $m_{1/2}$,
which is crucial. Both solutions increase with $\tan\beta$, as
expected by the bottom Yukawa dependence explicit in
eq.~(\ref{gtn2coshb}) and remain practically constant for large
$m_{1/2}$ values.

\section{Two-Body Decays of the Lightest Stop: the \rp violating  case}

In contrast to the case of an R--parity conserving supergravity
theory, in our broken R--parity model one can have a competing \rp
violating stop decay mode in region I of Fig.~\ref{par-sp}. From
eq.~(\ref{sqtocha}) with $\tau=F_3^+$ one can easily compute the \rp
violating stop decay width $\tilde t_1 \to b\,\tau$,
\begin{eqnarray}
\Gamma(\s t_1 \to b\,\tau)&=&
\frac{g^2\lambda^{1/2}(m_{\s t_1}^2,m_{b}^2,m_{\tau}^2)}{
16\pi m_{\s t_1}^3}\{-4U_{32}^*\hat h_bc_{\theta_{\s t}}(V_{32}^*\hat
h_ts_{\theta_{\s t}}-V_{31}^*c_{\theta_{\s t}})m_{b}m_{\tau}\nn\\
&&+[(V_{32}^*
\hat h_ts_{\theta_{\s t}}-V_{31}^*c_{\theta_{\s t}})^2+U_{32}^{*2}
\hat h_b^2c^2_{\theta_{\s t}}](m_{\s t_1}^2-m_{b}^2-m_{\tau}^2)\}
\label{taue}
\end{eqnarray}
which coincides with the result found in Ref.~\cite{stop1}.
In~\cite{akeroyd} it was shown that, except for $U_{32}$ which
determines the SU(2)-conserving mixing of the Higgsino with the
left-handed $\tau$, all other mixing matrix elements $V_{3i}$ and
$U_{3i}$ are proportional to $v_3^{II}$ and therefore to the
tau--neutrino mass.  Neglecting these terms we have from
eq.~(\ref{taue})
\begin{equation}
\Gamma(\s t_1 \to b\,\tau)\approx
\frac{g^2\lambda^{1/2}(m_{\s t_1}^2,m_{b}^2,m_{\tau}^2)}{
16\pi m_{\s t_1}^3}\sin^2\xi'
\hat h_b^2c^2_{\theta_{\s t}}(m_{\s t_1}^2-m_{b}^2-m_{\tau}^2)
\label{tausa}
\end{equation}
noting that, to a good approximation,
\begin{equation}
|U_{32}|\approx\left|\frac{\epsilon_3}{\mu'}\right|=|\sin\xi'|
\label{u32}
\end{equation}
where $\epsilon_3$ corresponds to the bilinear mass parameter in basis
I.  The lesson here is that the R-Parity violating decay rate
$\Gamma(\s t_1 \to b\,\tau)$ is proportional to $\epsilon_3$ or,
equivalently, to $\sin^2\xi'$, instead of $\sin^2\xi$, and thus not
necessarily small, since it is not directly controlled by the neutrino
mass. In other words, there can be cancellations in the latter but not
in the \rp violating branching ratio.

The meaning of the factor $\sin^2\xi' h_b^2$ may also be seen 
in basis II, where $\epsilon^{II}_3=0$.  In this case $v_3^{II}$
is proportional to the tau--neutrino mass so that, as already
mentioned, in this basis all the elements $U_{3i}$ and $V_{3i}$ are
small~\cite{javi}.  Neglecting these terms, $\Gamma(\s t_1 \to
b\,\tau)$ may be written directly from the interaction term $\s
t_Lb_R\tau_L$, which is induced by the trilinear term in the
$\epsilon^{II}_3=0$--basis given in eq.~(\ref{supxip}) as
\begin{equation}
\lambda_3^{II}=\left(\epsilon_3/\mu'\right)h_b=h_b\sin\xi'
\end{equation}
which is the factor in eq.~(\ref{tausa}). Note, however, that in our
numerical calculation to be described in the next section we have used
for $\Gamma(\s t_1 \to b\,\tau)$ the full expression given in
eq.~(\ref{tause}) of the appendix.

In the next section we will determine the conditions under which the
\rp violating decay width $\Gamma(\s t_1 \to b\,\tau)$ can be dominant
over the \rp conserving ones, $\Gamma(\s t_1 \to c\,\s\chi_1^0)$ and
$\Gamma(\s t_1 \to b\,\s\chi_1^+)$.

\subsection{Region I}

\begin{figure}
\centerline{\protect\hbox{
\psfig{file=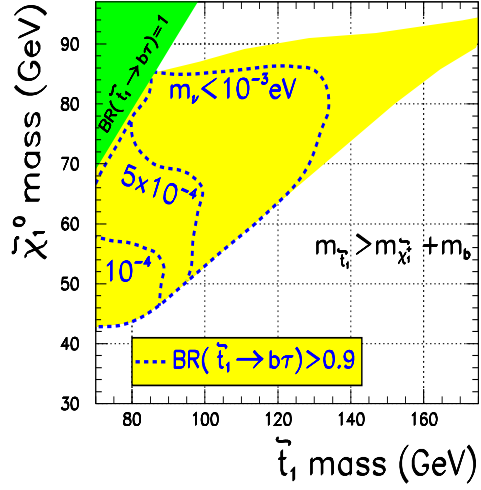,height=7.0cm,width=7.5cm}}}
\caption{\small   
  Regions where the ${\tilde t}_1 \to b\,\tau$ decay branching ratio
  exceeds 90\% in the $m_{{\tilde t}_1}$--$m_{{\tilde\chi}_1^0}$ plane
  for different \mnt values.  The MSSM parameters are randomly varied
  as indicated in the text under the restriction $m_{\tilde
    t_1}<m_{\s\chi^{\pm}_1}+m_b$.  The upper--left triangular region
  corresponds to $m_{\tilde t_1}<m_{\s\chi^0_1}+m_c$ so that only the
  $\tilde t_1 \to b\,\tau$ decay channel is open. The lower--right
%%%
  unshaded % this is the new
  region corresponds to $m_{\tilde
    t_1}>m_{\s\chi^+_1}+m_b$. }
\label{reg1-2}
\end{figure}

Using the one-step approximation for $\Gamma(\s t_1 \to c\,\s\chi_1^0)$
one finds
\begin{equation}
\Gamma(\s t_1 \to c\,\s\chi_1^0)\sim 10^{-6}h_b^4m_{\s t_1} 
\left(1-\frac{m_{\s \chi_1^0}^2}{m_{\s t_1}^2}\right)^2
\end{equation}
Using the eq.~(\ref{tausa}) and neglecting charm, tau and bottom
masses we get
\begin{equation}
\frac{\Gamma(\s t_1 \to c\,\s\chi_1^0)}{\Gamma(\s t_1 \to b\,\tau)}\sim
10^{-5} \frac{h_b^2}{\sin^2\xi'}
\left(1-\frac{m_{\s \chi_1^0}^2}{m_{\s t_1}^2}\right)^2
\end{equation}
Therefore $\Gamma(\s t_1 \to c\,\s\chi_1^0)$ will start to compete
with $\Gamma(\s t_1 \to b\,\tau)$ from $\sin\xi'\lsim 5\times 10^{-3}$
($10^{-4}$) for $\tan\beta$ large (small). In Fig. \ref{reg1-2} we
compare $BR({\tilde t}_1 \to c\,{\tilde{\chi}_1^0})$ [calculated
numerically from their exact formula in~(\ref{tause})] with
$BR({\tilde t}_1 \to b\,\tau)$ within the restricted region of the
$m_{{\tilde t}_1}$--$m_{{\tilde\chi}_1^0}$ plane where only those two
decay modes are open. We consider different \mnt values (these
correspond to relatively small values of the \rp parameters
$|\epsilon_3|, |v_3| \lsim 1$ GeV). We vary the MSSM parameters
randomly obeying the condition $m_{\tilde t_1}<m_{\s\chi^{\pm}_1}+m_b$
and depict the corresponding region in light grey. The upper--left
triangular region is defined by kinematics and corresponds to
$m_{\tilde t_1}<m_{\s\chi^0_1}+m_c$, so that $BR(\tilde t_1 \to
b\,\tau)=100 \: \%$. The lower--right grey corresponds to $m_{\tilde
  t_1}>m_{\s\chi^0_1}+m_c$ when the sampling is done over the region
defined by \eq{scan0}.  One notices from Fig.  \ref{reg1-2} that in
the central region the dominant stop decay mode is $\tilde t_1 \to
b\,\tau$ with branching ratio $BR(\tilde t_1 \to b\, \tau)>0.9$.  The
dotted lines in the light grey region indicate maximum \nt mass values
obtained in the scan. In the calculation of the \nt mass, we have
allowed only up to one order of magnitude of cancellation between the
two terms which contribute to $\sin\xi$.  Therefore if the lightest
stop only decays into the two modes considered here, the processes $\s
t_1 \to b\,\tau$, will be important even for the case of very light
tau--neutrino masses.

We note however that we can use the limits obtained from leptoquark
searches~\cite{Abe:1997dn} in order to derive limits on the top-squark
for our \rp violating case. In particular, if $BR({\tilde t}_1\to
b\tau)=1$ stop masses less than 99 GeV are excluded at 95\% of CL.,
under the assumption that the three--body decays of the stops are
negligible.  Therefore, the dark region in Fig.~\ref{reg1-2} would be
ruled out.  In ref.~\cite{deCampos:1998kw} we have determined the
corresponding restrictions on the SUGRA parameter space.

\begin{figure}
\centerline{\protect\hbox{
\psfig{file=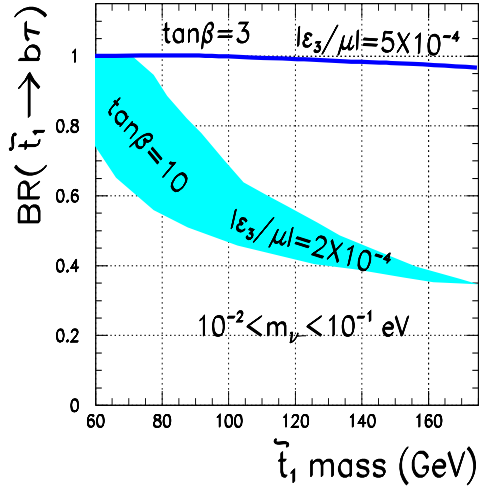,height=7.0cm,width=7.5cm}}}
\caption{\small  $BR(\s t_1 \to b\,\tau)$ as function of
  the lighter stop mass for tau--neutrino mass in the sub--eV range
  and two different values of $\tan\beta$ and $\epsilon_3/\mu$.  This
  prediction is natural in the sense that we have allowed only up to
  one order of magnitude of cancellation between the two terms that
  contribute to $\sin\xi$.  }
\label{bstbtatm}
\end{figure}

The dependence on the tau neutrino mass may be seen in
Fig.~\ref{bstbtatm} where the role played by $\tan\beta$ is manifest.
In this figure we have shown $BR(\s t_1 \to b\,\tau)$ as function of
the lighter stop mass for tau--neutrino mass in the sub--eV range,
indicated by the simplest oscillation interpretation of the
Super-Kamiokande atmospheric neutrino data.
We have obtained such tau--neutrino mass values numerically, allowing
only one decade of cancellation between the two terms that contribute
to $\sin\xi$ in \eq{sinxi}. The degree of suppression for
$\epsilon_3/\mu$ obtained numerically agrees very well with the
expectations from the approximate formula for the minimal
tau--neutrino mass in eq.~(\ref{mNeutrinods}). In contrast with
Ref.~\cite{stop1}, in our case $BR({\tilde t}_1\to b\tau)$ decreases
with $\tan\beta$. The reason for this difference is that here we take
into account the fact that the mixing parameter $\Gamma_{UL13}$
obtained from the RGE depends on $h_b^2$ in eq.~(\ref{gtn2cos}), while
in ref.~\cite{stop1} was simply regarded as a phenomenological input
parameter (called $\delta$ there).

The message from this subsection is that in our SUGRA \rp violating
model the R-Parity violating decay mode $\s t_1 \to b\,\tau$ can very
easily dominate the R-Parity conserving decay mode $\s t_1 \to
c\,\s\chi_1^0$, even for very small neutrino masses.

\subsection{Region II}

In region II the R--Parity conserving decay mode $\tilde t_1 \to b
\tilde\chi^+_1$ is open (but not $\tilde t_1 \to t \nu$), and competes
with the R--Parity violating mode $\tilde t_1 \to b\,\tau$.  Replacing
the subindex 3 by 1 on the diagonalization matrices $U$ and $V$ in
eq.~(\ref{taue}) we get the corresponding expression for $\Gamma(\s
t_1 \to b\,\s\chi_1^+)$. In order to get an approximate expression for
the ratio of the two main decay rates in this region, we note that in
MSUGRA the lightest chargino is usually gaugino-like, implying that
$V_{11}^2\sim 1$. In addition, the lightest stop is usually
right-handed, hence $\sin^2\theta_{\s t}\gsim\cos^2\theta_{\s t}$.
This way we find
\begin{equation}
\frac{\Gamma(\tau)}{\Gamma(\s\chi_1^+)} \equiv
\frac{\Gamma(\s t_1 \to b\,\tau)}{\Gamma(\s t_1 \to
b\,\s\chi_1^+)}\approx
\frac{\sin^2\xi'\hat h_b^2\cos^2{\theta_{\s t}}}
{\left[(V_{11}^{*}\cos{\theta_{\s t}}-
V_{12}^{*}\hat h_t\sin{\theta_{\s t}})^2+U_{12}^{*2}\hat h_b^2
\cos^2{\theta_{\s t}}\right]}\,K
\label{Gtc}
\end{equation}
where $K$ is a kinematical factor depending on the lightest stop and
chargino masses, and here we have defined $\hat h_{t,b} \equiv
h_{t,b}/g$. The presence of the bottom quark Yukawa coupling indicates
that large values of $\tan\beta$ are necessary to have large R--Parity
violating branching ratios in this region. In fact, we have checked
numerically with the exact expressions that in Region II (RII)
${\Gamma(\tau)}/{\Gamma(\s\chi_1^+)}\gsim1$ only for large $\tan\beta$
as we will see in the next figures. 

In Fig.~\ref{reg2} we show the regions in the 
$m_{\chi^{\pm}_1}-m_{\tilde t_1}$ plane where $BR(\tilde t_1 \to
b\,\tau)$ 
dominates over $BR(\tilde t_1 \to b\widetilde\chi^+_1)$. In the
upper--left 
region the decay mode $\tilde t_1 \to b\widetilde\chi^+_1$ is not
allowed 
and corresponds to Region I. Below and to the right of this zone, and
above 
and to the left of three rising lines, lies region RII where
${\Gamma(\tau)}/{\Gamma(\s\chi_1^+)}>1$. The three lines correspond to
$|\epsilon_3|<80$ GeV (dashed), $|\epsilon_3|<60$ GeV (dotted), and
$|\epsilon_3|<40$ GeV (dot--dashed), respectively. The proximity to
the upper-left zone indicates that the RPV decay dominates only close
to the threshold where there is a high kinematical suppression of the
R--parity-conserving one, through the factor $K$. Unlike the case of
region I this requires large values of the RPV parameters.  Note,
moreover, that if the stops have a small mixing ($\cos{\theta_{\s
    t}}\approx 0$), then ${\Gamma(\tau)}/{\Gamma(\s\chi_1^+)} \ll 1$
in RII.
\begin{figure}
\centerline{\protect\hbox{
\psfig{file=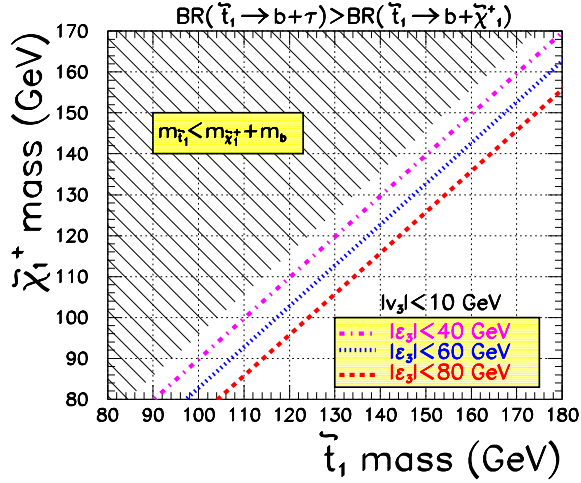,height=7cm}}}
\caption{\small 
  Contours of $BR({\tilde t}_1 \to b\,\tau)>BR({\tilde t}_1 \to
  b\s\chi_1^+)$ in the $m_{{\tilde t}_1}$--$m_{{\s\chi}_1^+}$ plane
  for $|v_3|<10$ GeV.  Three different maximum values for
  $|\epsilon_3|$ are considered: $|\epsilon_3|<40$ GeV (dot-dash),
  $|\epsilon_3|<60$ GeV (dots), and $|\epsilon_3|<80$ GeV (dashes).
  The region where $m_{\tilde t_1}<m_b+m_{\widetilde\chi^+_1}$
  corresponds to the previously studied Region I.}
\label{reg2}
\end{figure}

A simpler expression for the ratio of decay rates in eq.~(\ref{Gtc})
is obtained if we take $V_{11}\approx1$ and assume no kinematical
suppression in eq.~(\ref{Gtc}) through the factor $K$:
\begin{equation}
\frac{\Gamma(\tau)}{\Gamma(\s\chi_1^+)}\sim
\sin^2\xi'\hat h_b^2\,.
\label{Gta}
\end{equation}
Note that the presence of the parameter $\sin\xi'=\epsilon_3/\mu'$
indicates that the R--Parity violating decay mode is \emph{not}
strictly proportional to the neutrino mass, but proportional to the
BRpV parameter $\epsilon_3^2$. 

\begin{figure}
\centerline{\protect\hbox{
\psfig{file=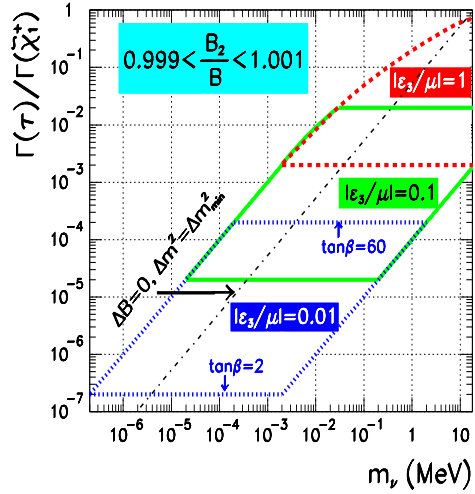,height=7.0cm,width=7.5cm}}}
\caption{\small
  Regions for $\Gamma({\tilde t}_1 \to b\,\tau)/\Gamma({\tilde t}_1
  \to b\, {\tilde{\chi}_1^+})$ as a function of the tau--neutrino mass
  with the universality condition $B=B_3$ at the unification scale
  imposed at the 0.1\% level as indicated. Its effect is to alter the
  maximum attainable tau--neutrino mass. The dot-dashed line
  corresponds to the case where $\Delta B=0$ at the weak scale.}
\label{nueuni}
\end{figure}

However generically we expect some correlation with the \nt mass,
especially in the case where the boundary conditions in the RGE are
universal and there are no strong cancellations between two terms that
contribute to $\sin\xi$ as shown in Fig.~\ref{nueuni}.  In this figure
we plot the ratio $\Gamma(\tau)/\Gamma(\s\chi_1^+)$ in RII as a
function of the tau--neutrino mass. Both decay rates have been
calculated numerically from the exact formulas. In this figure we have
imposed both $m_{H_1}^2=M_L^2$ and $B=B_3$ within 0.1\% at the GUT
scale.  Cancellation between the $\Delta m^2$ and $\Delta B$ terms in
the neutrino mass formula of eq.~(\ref{sinxi}) are accepted only
within 1 decade. As a reference we have drawn the line corresponding
to $\Delta B=0$ and $\Delta m^2=\Delta m^2_{min}$ ($\Delta m^2$ is
negative and its magnitude is bounded from below by $\Delta
m^2_{min}$) at the weak scale, which gives an idea of the value of the
neutrino mass when there is no cancellation between the $\Delta B$ and
$\Delta m^2$ terms.

We have imposed an upper bound on $m_{\nu_{\tau}}$ at the collider 
experimental limit of the tau--neutrino mass, and have chosen fixed
values 
of $\epsilon_3/\mu=1$, 0.1, and 0.01.  The allowed region for
$\epsilon_3/\mu=1$ is above the dashed line.  In the case of
$\epsilon_3/\mu=0.1$ (0.01) the allowed region lies enclosed between
the solid (dotted) lines. The effect of $\tan\beta$ is to increase the
ratio $\Gamma(\tau)/\Gamma(\s\chi_1^+)$: the minimum value of the
ratio is obtained for $\tan\beta\approx 2$ and the maximum corresponds
to $\tan\beta \approx60$.  The extreme values of $\tan\beta$ are
dictated by perturbativity.

%%%%%%%%%%%%%%%%%%%%%%%%%%%%%%%%%%%%%%
%%%%%%%%%%%%%%%%%%%%%%%%%%%%%%%%%%%%%%%%%

A number of statistically less significant points appear outside the
drawn regions in Fig.~\ref{nueuni} and are not depicted. They
correspond to points with $m_{\tilde
  t_1}-m_b-m_{{\tilde\chi}_1^\pm}<10\,$ GeV which appear above the
diagonal line, and points with $\cos\theta_{\tilde t}<0.1$ which
appear below the horizontal line corresponding to the lowest values of
$\tan\beta$. In the last case, our approximation in eq.~(\ref{Gta})
does not work any more. On the other hand, eq.~(\ref{Gta}) predicts
very well the behavior of $\Gamma(\tau)/\Gamma(\s\chi_1^+)$ if
$\cos\theta_{\s t}>0.1$. For example for $\epsilon_3/\mu=1$, or
equivalently $\sin\xi'=1/\sqrt2$, we expect from eq.~(\ref{Gta}) a
maximum value of order 1 for large $\tan\beta$ ($h_b\approx1$) and a
minimum value of order $10^{-3}$ for small $\tan\beta$ ($h_b^2\approx
10^{-3}$), and this is confirmed by Fig.~\ref{nueuni}. High values of
the R--parity violating branching ratio for large $\epsilon_3$ values
are highly restricted for large $\tan\beta$. This can be understood as
follows.  In the case of $\epsilon_3/\mu=1$ and $\tan\beta=60$
acceptable neutrino masses are obtained only if $\sin\xi\sim 1$. On
the other hand, in this regime we find from eq.~(\ref{sinxi}) that the
$\Delta B$ term is large because of the high value of $\tan\beta$, and
that the $\Delta m^2$ term is large because $m_{H_1}^2$ becomes
negative and $\Delta m^2=m_{H_1}^2-M_L^2$ grows in magnitude. This
way, acceptable neutrino masses are achieved only with cancellation
within more than one decade. In any case, we think that
Fig.~\ref{nueuni} is very conservative considering that in MSSM--SUGRA
with unification of top-bottom-tau Yukawa couplings, the large value
of $\tan\beta$ implies that a cancellation of four decades among vev's
is needed.

%%%%%%%%%%%%%%%%%%%%%%%%%%
\begin{figure}
\centerline{\protect\hbox{
\psfig{file=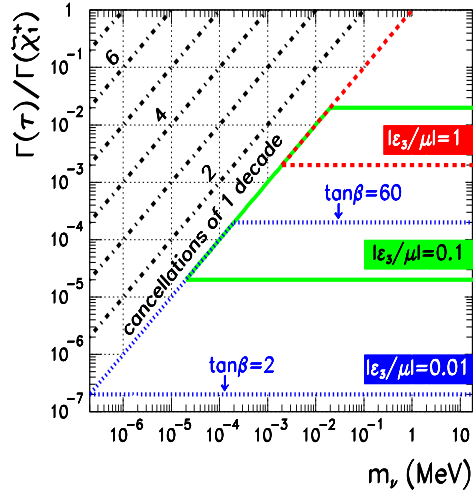,height=7.0cm,width=7.5cm}}}
\caption{\small
  Regions for $\Gamma({\tilde t}_1 \to b\,\tau)/\Gamma({\tilde t}_1
  \to b\, {\tilde{\chi}_1^+})$ as a function of the tau--neutrino mass
  for different levels of cancellation between the two terms that
  contribute to the neutrino mass. We impose the universality
  condition $m_{H_1}^2=M_L^2$ at the unification scale, but $B$ and
  $B_3$ are not universal. We take $|\epsilon_3/\mu|=1$ (inside the
  dashed lines), $|\epsilon_3/\mu|=0.1$ (solid lines), and
  $|\epsilon_3/\mu|=0.01$ (dotted lines).}
\label{Gtaumnu}
\end{figure}

The width of the band in Fig.~\ref{nueuni} reflects the degree of
correlation between the ratio $\Gamma({\tilde t}_1 \to b\,\tau)/\Gamma
({\tilde t}_1 \to b\, {\tilde{\chi}_1^+})$ and the neutrino mass under
the mentioned conditions. Note that one would have an indirect
measurement of the neutrino mass if this ratio were determined
independently. The band will open to the left if one allows a stronger
cancellation between the terms in $\Delta B$ and $\Delta m^2$. On the
other hand it will open to the right if the universality between $B$
and $B_3$ is relaxed. This is shown in Fig.~\ref{Gtaumnu} where we
plot the ratio $\Gamma(\tau)/\Gamma(\s\chi_1^+)$ in RII as a function
of the tau--neutrino mass, but without imposing universality between
$B_3$ and $B$. If we accept cancellation within one decade between the
$\Delta B$ and $\Delta m^2$ terms , then the allowed region is at the
right and below the corresponding dashed tilted line.  If a larger
degree of cancellation is accepted, the left boundary of the allowed
region moves to the left as indicated in the figure, enhancing the \rp
violating channel.  In addition if we accept only a decade of
cancellation between the two terms that contribute to the
tau--neutrino mass, then our approximate formula which predicts the
minimum tau--neutrino mass in eq.~(\ref{mNeutrinods}) works very well.

\begin{figure}
\centerline{\protect\hbox{
\psfig{file=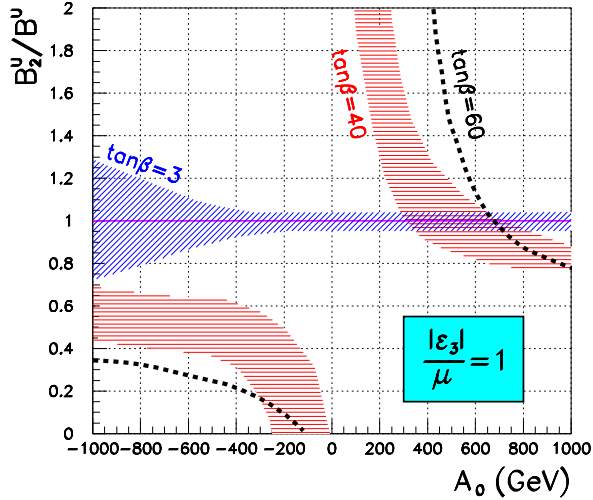,height=7cm}}}
\caption{\small 
  Universality condition $B=B_3$ at the unification scale as a
  function of $A_0$. As $\tan\beta$ increases, the allowed values of
  $A_0$ are more constrained.}
\label{b2ubu}
\end{figure}

In summary, in this subsection we have shown that even in region II,
where the R-Parity conserving decay mode ${\tilde t}_1 \to
b\,{\tilde{\chi}_1^+}$ is also open, the R-Parity violating decay mode
${\tilde t}_1\to b\,\tau$ can be comparable to ${\tilde t}_1\to
b\,{\tilde{\chi}_1^+}$ for large $\tan\beta$ and $\epsilon_3$, and
relatively close to the chargino production threshold. In general,
this implies a large neutrino mass unless a cancellation is accepted
between the two terms contributing to the tree level neutrino mass. In
addition, the non-universality of the $B$ and $B_3$ terms at the GUT
scale does not increase appreciably the allowed parameter space,
except at large $\tan\beta$. The main consequence of this
non-universality is to restrict the allowed values of $A_0$ at large
$\tan\beta$.  In the next subsection we study the effects introduced
by the non-universality of $m_{H_1}^2$ and $m_{L_3}^2$.

\subsection{Effects of non--Universality}

We now study the effect of possible non-universality of soft-breaking
SUSY parameters on our previous results. In particular, the
non-universality between $m_{H_1}^2$ and $m_{L_3}^2$ at the GUT scale.
The Minimal SUGRA model, while highly predictive, rests upon a number
of simplifying assumptions which do not necessarily hold in specific
models due to the possible evolution of the physical parameters in the
range from $M_{Planck}$ to $M_{GUT}$. Specifically, there are several
models in the literature with non-universal soft SUSY breaking mass
parameters at high scales.  A recent survey can be found in
\cite{nusug}, where several models such as based on string theory,
M-theory, and anomaly mediated supersymmetry are analyzed. For this
reason we find interesting to explore here the effects of
non-universal soft terms.

The SUGRA spectra are typically found for given values of $m_{1/2}$,
$m_0$, $A_0$, $\tan\beta$ and Sgn($\mu$). In our case we have in
addition $\sin\xi'$ (or equivalently, $\epsilon_3$). The value of
$v_3$ is determined by the previous parameters through the
minimization conditions.  In addition, a relation between $A_0$ and
the ratio $B_3/B$ at the GUT scale (which indicates the degree of
universality) emerges. This relation can be seen in Fig.~\ref{b2ubu}
for $\epsilon_3/\mu=1$ and the values $\tan\beta=3$, 40, and 60, for
$m_{H_1}^2=M_L^2$. The relation becomes more restrictive as
$\tan\beta$ is increased, starting from $-1000<A_0<1000$ GeV allowed
for $\tan\beta=3$, to a single $A_0$ value compatible with unification
for $\tan\beta=60$.
\begin{figure}
\centerline{\protect\hbox{
\psfig{file=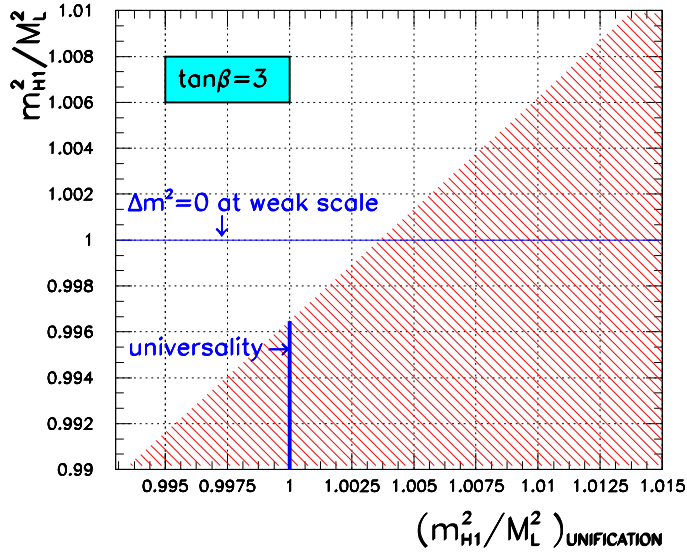,height=8.0cm}}}
\caption{\small Comparison between the ratio $m_{H_1}^2/M_L^2$ at the
  weak and the unification scales for $\tan\beta=3$. Universality at
  the unification scale, $m_{H_1}^2/M_L^2=1$, implies a maximum value
  for this ratio at the weak scale.}
\label{Dm2uni}
\end{figure}

Another way to enhance the \rp violating channel, enlarging the band
towards the left in Fig.~\ref{nueuni}, is by relaxing the universality
between $m_{H_1}^2$ and $M_L^2$ at the GUT scale. In Fig.~\ref{Dm2uni}
we plot the ratio $m_{H_1}^2/M_L^2$ at the weak scale as a function of
the same ratio at the unification scale $M_{GUT}$ for $\tan\beta=3$.
The shaded region is allowed, implying a maximum value for the ratio
$m_{H_1}^2/M_L^2$ at the weak scale for a given value of the ratio at
the GUT scale. We see from Fig.~\ref{Dm2uni} that a relaxation of
universality of 0.5\% or more is enough to make
$(m_{H_1}^2/M_L^2)_{weak}=1$ possible, meaning that smaller neutrino
masses are attainable without having to rely on a cancellation between
the $\Delta m^2$ and $\Delta B$ terms or small values of
$\Gamma({\tilde t}_1\to b\tau)$.
\begin{figure}
\centerline{\protect\hbox{
\psfig{file=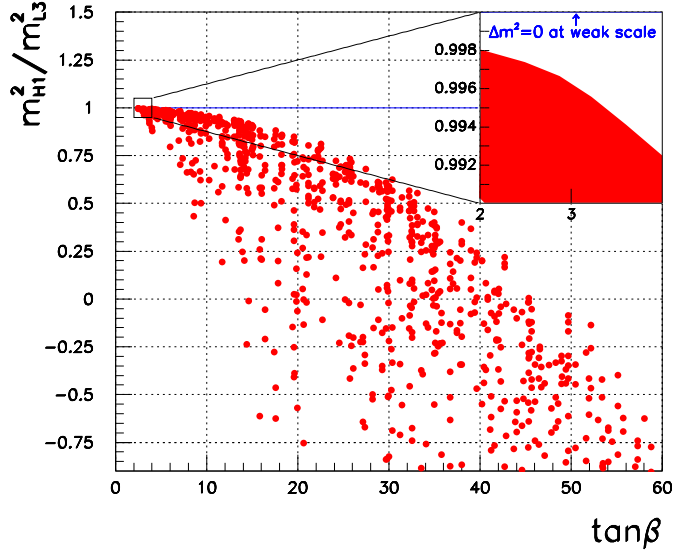,height=8.0cm}}}
\caption{\small
  $(m_{H_1}^2/M_L^2)$ evaluated at the weak scale as a function of
  $\tan\beta$. This ratio is always less than one and decreases with
  $\tan\beta$.}
\label{Dm2unitgb}
\end{figure}
However as we increase $\tan\beta$ the maximum value of
$m_{H_1}^2/M_L^2$ decreases, and thus, the required non-universality
between $m_{H_1}$ and $M_L$ at unification scale grows drastically. In
Fig.~\ref{Dm2unitgb} we show the ratio $m_{H_1}^2/M_L^2$ at the weak
scale as a function of $\tan\beta$. We appreciate clearly the growing
of $|\Delta m^2|_{min}$ with $\tan\beta$. We remind the reader that
this kind of non--universality in the soft terms is not uncommon in
string models~\cite{BIM}, or GUT models based on
$SU(5)$~\cite{Pomarol} or $SO(10)$~\cite{so10} for example.  
There are in fact some $SO(10)$ models for non-universality of the GUT
scale scalar masses which naturally favour light neutrino mass
\cite{so10}.

\begin{figure}
\centerline{\protect\hbox{
\psfig{file=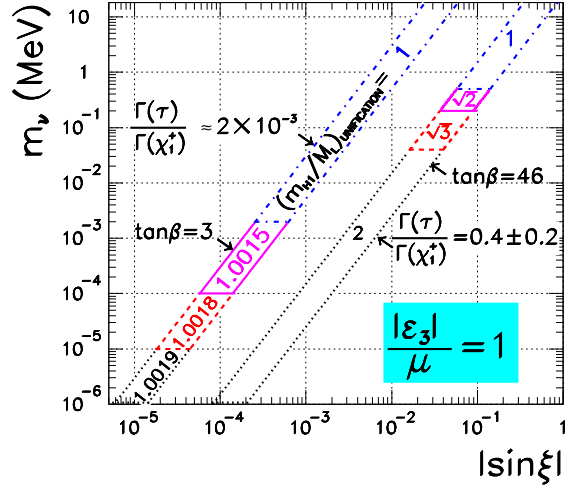,height=7cm}}}
\caption{\small 
  Minimum value of the tau--neutrino mass as a function of $\sin\xi$
  for different values of $m_{H_1}/M_L$ at the GUT scale and two
  values of $\tan\beta$. The ratio $\epsilon_3/\mu$ is fixed to the
  indicated value, leading to a nearly constant value for
  $\Gamma({\tilde t}_1 \to b\, \tau)/\Gamma({\tilde t}_1 \to
  b\,{\tilde{\chi}_1^+})$. Here we assume that the two terms
  contributing to the tau--neutrino mass cancel to within an order of
  magnitude.}
\label{mnu-sinxi}
\end{figure}
The effect of non--universality it is also explored in
Fig.~\ref{mnu-sinxi} where it is shown the relation between the
neutrino mass and the parameter $\sin\xi$ for $\epsilon_3/\mu=1$. Two
different bands are shown: one for $\tan\beta=3$ and
$\Gamma(\tau)/\Gamma(\s\chi_1^+)=2\times10^{-3}$, and a second one for
$\tan\beta=46$ and $\Gamma(\tau)/\Gamma(\s\chi_1^+)=0.4\pm0.2$. The
required degree of universality at the GUT scale is indicated inside
the bands.
For example, in order to have neutrino masses of the order of eV for
$\tan\beta=3$, $m_{H_1}^2$ needs to be at least 0.2\% larger than
$M_L^2$. Similarly, for $\tan\beta=46$ we need a $m_{H_1}^2$ twice as
large as $M_L^2$ at the GUT scale in order to have neutrino masses of
1 eV. We stress the fact that for Fig.~\ref{mnu-sinxi} we have
conservatively accepted cancellation at the level of one
order-of-magnitude only.

In summary, the lesson to learn here is that non-universal soft SUSY
breaking terms at the GUT scale have the potential of making it easier
to reconcile sizeable R-Parity violating effects in the stop sector
with very small neutrino masses, without resorting to cancellations.

\section{Conclusions}

We have studied the decays of the lightest top squark in SUGRA models
with and without R-parity. We have improved the calculation for the
decay $\s t_1 \to c\,\s\chi^0$ by numerically solving the
renormalization group equations (RGE's) of the MSSM including full
generation mixing in the RGE's for Yukawa couplings as well as soft
SUSY breaking parameters. The decay-width is in general one order of
magnitude smaller than the one obtained in the usual one--step
approximation. This result will therefore enlarge the regions of
parameter space where the four--body decays of lightest stop dominate
over the decay into a charm quark and the lightest neutralino. As a
result it will affect the present experimental lower bound on the $\s
t_1$ mass even in the R-parity conserving case~\cite{Boehm:1999tr}. If
R--parity breaks of course new decay modes appear and, as we have
shown, they can be sizeable. In fact we have shown that the lightest
stop can be the LSP, decaying with 100\% rate into a bottom quark and
a tau lepton. We have shown that the decay mode $\s t_1 \to b\, \tau$
dominates over $\s t_1 \to c\,\s\chi^0$ even for neutrino masses in
the range suggested by the simplest oscillation interpretation of the
Super-Kamiokande atmospheric neutrino data. This result would have a
strong impact on the top squark search strategies at
LEP~\cite{Abbiendi:1999yz} and TEVATRON~\cite{Holck:1999tp}, where it
is usually assumed that the $\s t_1 \to c\,\s\chi^0$ decay mode is the
main channel. In addition to the signal of two jets and two taus
present when the two produced stops decays through the R-parity
violating channel, one expects a plethora of exotic high--multiplicity
fermion events arising from neutralino decay, since such decay can
happen inside the detector even for the small neutrino masses in the
range suggested by the $\nu_\mu$ to $\nu_\tau$ oscillation
interpretation of the atmospheric neutrino
anomaly~\cite{Romao:1999up}. We have also compared the decay $\s t_1
\to b\,\tau$ with the R-parity and flavor conserving mode $\s t_1 \to
b\,\s\chi^+$ and shown that the rate of the former can be comparable
or even bigger than the latter if the tau--neutrino mass and
$\tan\beta$ are large.  However one may have a sizeable branching of
$\s t_1 \to b\,\tau$ in the case of suppressed tree level neutrino
mass as a result of strong cancellations between the two terms that
contribute to $\sin\xi$, or in some regions of parameter space of
non-universal SUGRA models with $(m^2_{H_1}/m^2_{L_3})_{GUT} \neq 1$. A
detailed analysis of the detectability prospects of such related
signatures at present and future accelerators lies outside of the
scope of the present paper and it will be taken up elsewhere.

\section*{Acknowledgments}

We thank J. Ferrandis, O. J. P. Eboli and W. Porod for useful
discussions.  This work was supported by DGICYT under grants PB95-1077
and by the TMR network grant ERBFMRXCT960090 of the European Union.
M.A.D. was supported by a DGICYT postdoctoral grant, by the U.S.
Department of Energy under contract number DE-FG02-97ER41022, and by
CONICYT grant 1000539. D.R. was supported by Colombian COLCIENCIAS
fellowship.

\appendix
%%%%%%%%%%%%%%%%%%%%%%%%%%%%%%%%%%%%%%%%%%%%%%%%%%%%%%%%%%%%%%%%%
%Please comment the following three lines if you like to use the usual
%format for appendix. Also change \sectionapp by \section
\renewcommand{\theequation}{\thesection.\arabic{equation}}
\def\sectionapp#1{\setcounter{equation}{0}
\section{#1}}
%%%%%%%%%%%%%%%%%%%%%%%%%%%%%%%%%%%%%%%%%%%%%%%%%%%%%%%%%%%%%%%%%

\sectionapp{Feynman Rules}
\label{appendixb}
In this Appendix we derive the Feynman rules $F_j^0q_i\tilde q_k$
(involving a neutralino/tau-neutrino, a quark, and a squark) and
$F_j^{\pm}q_i\tilde q'_k$ (involving a chargino/tau, a quark, and a
squark of different electric charge) in the case of three generations
and RPV in the third generation. This is a generalization of the
Feynman rules contained in~\cite{GH}, which are done for the R--Parity
conserving MSSM and for one generation of quarks and squarks.

Following~\cite{BBMR} we work in a quark interaction basis where 
$d_{L,R}=d_{L,R}^0$, $u_{L}=K u_{L}^0$, and $u_{R}=u_{R}^0$ (we 
denote $q$ and $q^0$ the mass and current eigenstates respectively), 
as opposed to Ref.~\cite{Rosiek} where a more general basis is used. 
In addition, we implement the notation $\tilde q_{L,R}\equiv \tilde 
q^0_{L,R}$ for the interaction basis.

The starting point is the following piece of the Lagrangian
\begin{eqnarray}
{\cal L}_{u \tilde u F^0}&=& 
-g{\bar u}_i^0
\Bigg\{
\sqrt2
\left[
\sin\theta_W e_U {N'}_{J1}+\frac 1{\cos\theta_W}
(\half-e_U\sin^2\theta_W){N'}_{J2}
\right]
\tilde u_{iL}\nonumber\\
&&+
\frac{m^{0U}_{ij}}{\sqrt2m_W\sin\beta\sin\theta}{N'}_{J4}\tilde u_{jR}
\Bigg\}
P_R F_J^0\nonumber\\
&&+g{\bar u}^0_i
\Bigg\{
\sqrt2
\left[
\sin\theta_W e_U {N'}_{J1}^*+\frac 1{\cos\theta_W}(-e_U\sin^2\theta_W)
{N'}_{J2}^*
\right]
\tilde u_{iR}\nonumber\\
&&-
\frac{m^{0U\dagger}_{ij}}{\sqrt2m_W\sin\beta\sin\theta}{N'}_{J4}^*\tilde 
u_{jL}\Bigg\}P_L F_J^0+\textrm{h.c.}
\label{nusu}
\end{eqnarray}
written in the quark interaction basis. The $5\times5$ matrix $N'$ 
diagonalizes the neutralino/neutrino mass matrix in the 
$(\tilde\gamma,\tilde Z,\tilde H_1^0,\tilde H_2^0,\nu_{\tau})$ basis 
as defined in~\cite{v3cha}, with the index $J=1...5$. The $3\times3$ 
up--type quark mass matrix $m^{0U}$ is not diagonal, with the indexes 
$i,j=1,2,3$. 

In order to write the above Lagrangian with mass eigenstates we use
the basic relations mentioned before, in particular, $u_{iL}^0=
(K^\dagger)^{ij}u_{jL}$, which implies that ${\bar u}_{iL}^0={\bar
u}_{jL}K^{ji}$. We need the following relations:
\begin{eqnarray}
{\bar u}_{iL}^0\tilde u_{iL}&=&
{\bar u}_{iL}\left(\Gamma_{UL}^{*}{(K^\dagger)}^*\right)^{ki}\tilde
u_{k}
\nonumber\\
{\bar u}_{iL}^0m^{0U}_{ij}\tilde u_{jR}
&=&{\bar u}_{iL}(\Gamma_{UR}^{*}m^U)^{ki}\tilde u_{k}
\nonumber\\
{\bar u}_{iR}\tilde u_{iR}&=&{\bar u}_{iR}\Gamma_{UR}^{*ki}\tilde u_{k}
\\
{\bar u}_{iR}m^{0U}_{ij}\tilde u_{jL}&=& 
{\bar u}_{iR}(\Gamma_{UL}^{*}K^*{m}^U)^{ki}\tilde u_{k}
\nonumber
\end{eqnarray}
where $i,j=1,2,3$ label the quark flavours, $k=1...6$ labels the
squarks, 
and $m^U\equiv{\mathrm{diag}}\,\{m_u,m_c,m_t\}$ is the diagonal up--type
quark mass matrix. In this way, the Lagrangian in eq.~(\ref{nusu}) can
be written as
\begin{equation}
{\cal L}_{u \tilde u F^0}= 
-g{\bar u}_i[(\sqrt2G_{0UL}^{*jki}+H_{0UR}^{*jki})P_R-
(\sqrt{2}G_{0UR}^{*jki}-H_{0UL}^{*jki})P_L]F_j^0\tilde u_k+\textrm{h.c.}
\label{luuN}
\end{equation}
where the different couplings are
\begin{eqnarray}
G_{0UL}^{jki}&=&\left[
\sin\theta_W e_U {N'}^*_{j1}+\frac 1{\cos\theta_W}
(\half-e_U\sin^2\theta_W){N'}^*_{j2}
\right]\left(\Gamma_{UL}K^\dagger\right)^{ki}\nonumber\\
G_{0UR}^{jki}&=&\left[
\sin\theta_W e_U {N'}_{j1}+\frac 1{\cos\theta_W}(-e_U\sin^2\theta_W)
{N'}_{j2}
\right]\Gamma_{UR}^{ki}\label{gh}\\
H_{0UL}^{jki}&=&{N'}_{j4}(\Gamma_{UL}K^{\dagger}{\hat h}_U)^{ki}
\nonumber\\
H_{0UR}^{jki}&=&{N'}^*_{j4}(\Gamma_{UR}\hat h_U)^{ki}\nonumber
\end{eqnarray}
and $\hat
h_U\equiv$diag$\,(m_u,m_c,m_t)/(\sqrt2m_W\sin\beta\sin\theta)$.
Graphically, the $F_j^0u_i\tilde u_k$ Feynman rules are given by

%%%%%%Neutralino--(u)quark--(u)squark%%%%%%%%%%%%%%%
\begin{picture}(120,60)(0,25) % y_2 for equation position
\DashArrowLine(60,25)(110,0){6}
\Vertex(60,25){3}
\ArrowLine(10,25)(60,25)
\ArrowLine(110,60)(60,25)
\Text(35,35)[]{$F_j^0$}
\Text(85,55)[]{$u_i$}
\Text(85,0)[]{${\tilde u}_k$}
\end{picture}
$
-ig[(\sqrt2G_{0UL}^{jki}+H_{0UR}^{jki})P_L
-(\sqrt{2}G_{0UR}^{jki}-H_{0UL}^{jki})P_R]
$

\begin{picture}(120,70)(0,25) % y_2 for equation position
\DashArrowLine(110,0)(60,25){6}
\Vertex(60,25){3}
\ArrowLine(60,25)(10,25)
\ArrowLine(60,25)(110,60)
\Text(35,35)[]{$F_j^0$}
\Text(85,55)[]{$u_i$}
\Text(85,0)[]{${\tilde u}_k$}
\end{picture}
$
-ig[(\sqrt2G_{0UL}^{*jki}+H_{0UR}^{*jki})P_R
-(\sqrt{2}G_{0UR}^{*jki}-H_{0UL}^{*jki})P_L]
$
\vspace{30pt} \hfill \\
The analogous Feynman rules in the MSSM are obtained by replacing
$F_i^0 \to \tilde\chi^0_i$, by interpreting the matrix $N$
as the usual $4\times4$ neutralino mixing matrix, and by setting
$\theta=\pi/2$ in the formula for the Yukawa couplings.

Similarly, replacing all $u(\tilde u)$ by $d(\tilde d)$ in 
eq.~(\ref{nusu}) and
starting from 
\begin{eqnarray}
{\cal L}_{q \tilde q' F^+}&=&g\bar{d}_i\left[
\frac{m^{0U\dagger}_{ij}}{\sqrt2m_W\sin\beta\sin\theta}V_{J2}\s
u_{jR}- V_{J1}\s
u_{iL}\right]P_RF^c_J
\nn\\
&&+g\bar{d}_i\frac{m^{D}_{ij}}{\sqrt2m_W\cos\beta\sin\theta}
U^*_{J2}\s u_{jL}P_LF^c_J\nn\\
&&+g\bar{u}^0_i\left[
\frac{m^{D\dagger}_{ij}}{\sqrt2m_W\cos\beta\sin\theta}U_{J2}\s
d_{jR}- U_{J1}\s
d_{iL}\right]P_RF^+_J
\nn\\
&&+g\bar{u}^0_i\frac{m^{0U}_{ij}}{\sqrt2m_W\cos\beta\sin\theta}
V^*_{J2}\s b_{jL}P_LF^+_J+\hbox{h.c}
\end{eqnarray}
we can obtain the complete Feynman
rules for the neutralino/tau--neutrino and chargino/tau with quarks and 
squarks. The results, that complements the obtained in~\cite{BBMR}, are

Neutralino--(d)quark--(d)squark
%%%%%%%%%%Neutralino--(d)quark--(d)squark%%%%%%%%%

\begin{picture}(120,60)(0,25) % y_2 for equation position
\DashArrowLine(60,25)(110,0){6}
\Vertex(60,25){3}
\ArrowLine(10,25)(60,25)
\ArrowLine(110,60)(60,25)
\Text(35,35)[]{$F_j^0$}
\Text(85,55)[]{$d_i$}
\Text(85,0)[]{${\tilde d}_k$}
\end{picture}
$
-ig[(\sqrt2G_{0DL}^{jki}+H_{0DR}^{jki})P_L
-(\sqrt{2}G_{0DR}^{jki}-H_{0DL}^{jki})P_R]
$

\begin{picture}(120,70)(0,25) % y_2 for equation position
\DashArrowLine(110,0)(60,25){6}
\Vertex(60,25){3}
\ArrowLine(60,25)(10,25)
\ArrowLine(60,25)(110,60)
\Text(35,35)[]{$F_j^0$}
\Text(85,55)[]{$d_i$}
\Text(85,0)[]{${\tilde d}_k$}
\end{picture}
$
-ig[(\sqrt2G_{0DL}^{*jki}+H_{0DR}^{*jki})P_R
-(\sqrt{2}G_{0DR}^{*jki}-H_{0DL}^{*jki})P_L]
$
\vspace{30pt} \hfill \\

\noindent
The mixing matrices $G_{0D}$ and $H_{0D}$ are defined as 
\begin{eqnarray}
G_{0DL}^{jki}&=&\left[
\sin\theta_W e_D {N'}_{j1}^*+\frac 1{\cos\theta_W}
(T_{3D}-e_D\sin^2\theta_W){N'}^*_{j2}
\right]\Gamma_{DL}^{ki}\nonumber\\
G_{0DR}^{jki}&=&\left[
\sin\theta_W e_D {N'}_{j1}+\frac 1{\cos\theta_W}(-e_D\sin^2\theta_W)
{N'}_{j2}
\right]\Gamma_{DR}^{ki}\nonumber\\
H_{0DL}^{jki}&=&{N'}_{j3}(\Gamma_{DL}{\hat h}_D)^{ki}\nonumber\\
H_{0DR}^{*jki}&=&{N'}_{j3}^*(\Gamma_{DR}\hat h_D)^{ki}
\end{eqnarray}

%\newpage

Chargino/tau--(d)quark--(u)squark

%%%%%% Chargino--(d)quark--(u)squark%%%%%%%%%%%%%%

\begin{picture}(120,60)(0,25) % y_2 for equation position
\DashArrowLine(60,25)(110,0){6}
\Vertex(60,25){3}
\ArrowLine(10,25)(60,25)
\ArrowLine(110,60)(60,25)
\Text(35,35)[]{$F_j^+$}
\Text(85,55)[]{$d_i$}
\Text(85,0)[]{${\tilde u}_k$}
\end{picture}
$
-ig(-C^{-1})[(G_{UL}^{jki}-H_{UR}^{jki})P_L-H_{UL}^{jki}P_R]
$

\vglue0.5truecm

\begin{picture}(120,70)(0,25) % y_2 for equation position
\DashArrowLine(110,0)(60,25){6}
\Vertex(60,25){3}
\ArrowLine(60,25)(10,25)
\ArrowLine(60,25)(110,60)
\Text(35,35)[]{$F_j^+$}
\Text(85,55)[]{$d_i$}
\Text(85,0)[]{${\tilde u}_k$}
\end{picture}
$
-ig[(G_{UL}^{*jki}-H_{UR}^{*jki})P_R-H_{UL}^{*jki}P_L]C
$
\vspace{30pt} \hfill \\

\noindent
where $C$ is the charge conjugation matrix (in spinor space) and the
mixing matrices $G_U$ and $H_U$ are defined as
\begin{eqnarray}
G_{UL}^{jki}&\equiv& V_{j1}^*\Gamma_{UL}^{ki},\qquad
H_{UL}^{jki}\equiv U_{j2}^*(\Gamma_{UL}\hat h_D)^{ki},\nonumber\\
H_{UR}^{jki}&\equiv&V_{j2}^*(\Gamma_{UR}\hat h_UK)^{ki},
\end{eqnarray}
Chargino/tau--(u)quark--(d)squark
%%%%%%%%%% chargino--(u)quark--(d)squark%%%%%%%%%

\begin{picture}(120,60)(0,25) % y_2 for equation position
\DashArrowLine(60,25)(110,0){6}
\Vertex(60,25){3}
\ArrowLine(10,25)(60,25)
\ArrowLine(110,60)(60,25)
\Text(35,35)[]{$F_j^+$}
\Text(85,55)[]{$u_i$}
\Text(85,0)[]{${\tilde d}_k$}
\end{picture}
$
-ig(-C^{-1})[(G_{DL}^{jki}-H_{DR}^{jki})P_L-H_{DL}^{jki}P_R]
$

\begin{picture}(120,70)(0,25) % y_2 for equation position
\DashArrowLine(110,0)(60,25){6}
\Vertex(60,25){3}
\ArrowLine(60,25)(10,25)
\ArrowLine(60,25)(110,60)
\Text(35,35)[]{$F_j^+$}
\Text(85,55)[]{$u_i$}
\Text(85,0)[]{${\tilde d}_k$}
\end{picture}
$
-ig[(G_{DL}^{*jki}-H_{DR}^{*jki})P_R-H_{DL}^{*jki}P_L]C
$
\vspace{30pt} \hfill \\

\noindent
where the mixing matrices $G_D$ and $H_D$ are defined as
\begin{eqnarray}
G_{DL}^{jki}&\equiv& U_{j1}^*(\Gamma_{DL}K^\dagger)^{ki},\qquad
H_{DL}^{jki}\equiv V_{j2}^*(\Gamma_{DL}K^\dagger\hat
h_U)^{ki},\nonumber\\
H_{DR}^{jki}&\equiv&U_{j2}^*(\Gamma_{DR}\hat h_DK^\dagger)^{ki},
\end{eqnarray}
In order to derive the decays widths we write, for example 
eq.~(\ref{luuN}) as
\begin{equation}
{\cal L}_{u \tilde u F^0}= 
g{\bar u}_i^0(f^{*jki}_UP_R
+h^{*jki}_UP_L)F_j^0\tilde u_k+\textrm{h.c}
\end{equation}
The result is
\begin{eqnarray}
\Gamma(\tilde q_k \to q_i+F_j^0)&=&
\frac{g^2\lambda^{1/2}(m_{\tilde q_k}^2,m_{q_i}^2,m_{F_j^0}^2)}{
16\pi m_{\tilde q_k}^3}\bigg[-4h^{jki}_Qf^{jki}_Q
m_{q_i}m_{F_j^0}
\nonumber\\
&&+\bigg((h^{jki}_Q)^2+(f^{jki}_Q)^2\bigg)\bigg(
m_{\tilde q_k}^2-m_{q_i}^2-m_{F_j^0}^2\bigg)\bigg]
\label{tause}\\
\Gamma(\tilde q_k \to q'_i+F_j^\pm)&=&
\frac{g^2\lambda^{1/2}(m_{\tilde q_k}^2,m_{q'_i}^2,m_{F_j^\pm}^2)}{
16\pi m_{\tilde q_k}^3}\bigg[-4l^{jki}_{Q}H^{jki}_{QL}
m_{q'_i}m_{F_j^\pm}
\nonumber\\
&&+\bigg((l^{jki}_{Q})^2+(H^{jki}_{QL})^2\bigg)\bigg(
m_{\tilde q_k}^2-m_{q'_i}^2-m_{F_j^\pm}^2\bigg)\bigg]
\label{sqtocha}
\end{eqnarray}
where $Q=U,D$ refers to $\tilde q$ and
\begin{eqnarray}
f^{jki}_Q&=&-(\sqrt2G_{0QL}^{jki}+H_{0QR}^{jki})\\
h^{jki}_Q&=&\sqrt{2}G_{0QR}^{jki}-H_{0QL}^{jki}\\
l^{jki}_Q&=&H_{QR}^{jki}-G_{QL}^{jki}
\end{eqnarray}
with the $G$ and $H$ couplings defined earlier in this appendix.

\end{document}